\documentclass[12pt]{article}
\usepackage{amsmath}
\usepackage{graphicx}
\usepackage{enumerate}
\usepackage{natbib}
\usepackage{url} 

\newcommand{\blind}{1}

\input{definition1}

\usepackage{epstopdf}

\begin{document}

\def\spacingset#1{\renewcommand{\baselinestretch}%
{#1}\small\normalsize} \spacingset{1}


\if1\blind
{
  \title{\bf Reducing multivariate independence testing to two bivariate means comparisons}
  \author{Kai  Xu\\
    School of Mathematics and Statistics, Anhui Normal University, \\
    Yeqing Zhou \\
    School of Mathematical Sciences, Tongji University,\\
    Liping Zhu\\
    Institute of Statistics and Big Data, 	Renmin University of China,\\
    and\\
    Runze Li\\
    Department of Statistics, The Pennsylvania State University.
}
  \maketitle
} \fi

\if0\blind
{
  \bigskip
  \bigskip
  \bigskip
  \begin{center}
    {\LARGE\bf  Reducing multivariate independence testing to two bivariate means comparisons
\end{center}
  \medskip
} \fi

\bigskip
\begin{abstract}
Testing for independence between two random vectors is a fundamental problem in statistics.
It is observed from empirical studies that many existing omnibus consistent tests may not work well for some strongly nonmonotonic and nonlinear relationships.
To explore the reasons behind this issue, we novelly transform the multivariate independence testing problem equivalently  into  checking the equality of two bivariate means.
An important observation we made is that the power loss is mainly due to cancellation of positive and negative terms in  dependence metrics, making them very close to zero.
Motivated by this observation, we propose a class of  consistent metrics with a positive integer $\gamma$ that exactly characterize independence.
Theoretically, we show that the metrics with even and infinity $\gamma$ can  effectively avoid the cancellation,
 and have high powers
under the alternatives that two mean differences offset each other.
Since we target at a wide range of dependence scenarios in practice, we further suggest to combine the p-values of test statistics with different $\gamma$'s through
the Fisher's method. We illustrate the advantages of our proposed tests  through   extensive numerical studies.

\end{abstract}

\noindent%

\spacingset{1.9} 
\section{Introduction}
\label{sec:intro}

 Testing for independence is a fundamental problem in statistics. It has received extensive attention over many decades.
In univariate scenarios, classic  measures of association such as Pearson's correlation \citep{Pearson:1895}, Kendall's $\tau$ \citep{Kendall:1938}, and Spearman's $\rho$ \citep{Spearman:1904} are widely used. However, a drawback of these measures is that they may be zero   when there is nonlinear association. In other words, tests  for independence based on these measures may lack  consistency against general types of alternatives. Alternative metrics, including Hoeffding's $D$ \citep{Hoeffding:1948}, Blum-Kiefer-Rosenblatt's $R$ \citep{Blum:Kiefer:Rosenblatt:1961}, and Bergsma-Dassios-Yanagimoto's $\tau^{\ast}$ \citep{Bergsma:Dassios:2014}, were developed to provide universally consistent guarantees against   general types  of dependence alternatives.
In many real-world applications, it is often necessary to test the independence between random vectors. 

To test the multivariate independence, various methods
	have been proposed in recent years.
A typical class of these methods is
to compare
the distance between  joint characteristic  or distribution  function of  random vectors  and the product of their marginal characteristic or distribution  functions.
Examples include
distance covariance \citep{Szekely:Rizzo:Bakirov:2007}, Hilbert-Schmidt independence criterion \citep{Grettonetal:2008}, Heller-Heller-Gorfine's coefficient \citep{Heller:Heller:Gorfine:2013}, projection covariance
\citep{Zhu:Xu:Li:Zhong:2017}, symmetric rank covariances \citep{Weihs:Drton:Meinshausen:2018},
multivariate Blum-Kiefer-Rosenblatt's and Bergsma-Dassios-Yanagimoto's coefficients \citep{Kim:Balakrishnan:Wasserman:2020a},
interpoint-ranking sign covariance \citep{Moon:Chen:2020},
ball covariance \citep{Panetal:2020},
center-outward rank-based dependence measures \citep{Shi:Drton:Han:2022,Shi:Hallin:Drton:Han:2022,Deb:Sen:2023} and among others.
Other types of multivariate tests include, but are not limited to,
graph-based
\citep{Berrett:Samworth:2019,Shen:Priebe:Vogelstein:2020,Azadkia:Chatterjee:2021},
ensembling \citep{Lee:Zhang:Kosorok:2022} and binning \citep{Gorsky:Ma:2022} methods.

We also remark that a problem closely related to testing independence is testing the equality of distributions.
Testing for independence reduces to a two-sample problem
via introducing an auxiliary label vector \citep{Heller:Heller:2016,Dhar:Dassios:Bergsma:2016, heller2016consistent}.
Conversely, however, it is still interesting to
	design the consistent independence tests originating from the perspective of two-sample testing.

In this article, we novelly transform the independence testing problem into checking the equality of  two bivariate means.
To elucidate the motivation behind this transformation, we first revisit the existing dependence metrics
under a unified framework: $\mu_{f_{1}, f_{2}}(\x, \y)\defby S_{1, f_{1}, f_{2}}+S_{2, f_{1}, f_{2}}-2S_{3, f_{1}, f_{2}}$, where $S_{l, f_{1}, f_{2}}$ for $l=1,2, 3$ denotes the
expectation of 	the product of kernel functions $f_{1}$ and $f_{2}$, using different independent copies of random vectors $\x \in \mR^{d_1}$ and $\y \in \mR^{d_2}$.
With different choices of functions $f_{1}$ and $f_{2}$, $\mu_{f_{1}, f_{2}}(\x, \y)$
encompasses many popular dependence measures. For example, if $f_{1}$ and $f_{2}$ are Euclidean distance, $S_{1, f_{1}, f_{2}}=E(\|\x_1-\x_2\|\|\y_1-\y_2\|)$,  $S_{2, f_{1}, f_{2}}=E(\|\x_1-\x_2\|\|\y_3-\y_4\|)$ and  $S_{3, f_{1}, f_{2}}=E(\|\x_1-\x_2\|\|\y_1-\y_3\|)$, then $\mu_{f_{1}, f_{2}}(\x, \y)$ is distance covariance \citep{Szekely:Rizzo:Bakirov:2007}.
We focus on the functions $f_{1}$ and $f_{2}$ that make $\mu_{f_{1}, f_{2}}(\x, \y)$ equal to zero if and only if the multivariate independence holds.
Based on this, we make an improtant observation: $S_{1, f_{1}, f_{2}}=S_{2, f_{1}, f_{2}}=S_{3, f_{1}, f_{2}}$ if and only if $\x$ and $\y$ are independent.
This finding has an important implication.
That is, the independence testing problem can be equivalently transformed into testing whether
 the bivariate mean vectors $(S_{1, f_{1}, f_{2}}, S_{3, f_{1}, f_{2}})$ and $(S_{3, f_{1}, f_{2}}, S_{2, f_{1}, f_{2}})$ are the same.

To test the null hypothesis $H_0': (S_{1, f_{1}, f_{2}}, S_{3, f_{1}, f_{2}})=(S_{3, f_{1}, f_{2}}, S_{2, f_{1}, f_{2}})$, we consider a class of metrics
with a positive integer $\gamma\in[1, \infty]$: $\mu_{\gamma,f_{1}, f_{2}}(\x, \y)\defby	\big\{(S_{1, f_{1}, f_{2}}-S_{3, f_{1}, f_{2}})^\gamma+(S_{2, f_{1}, f_{2}}-S_{3, f_{1}, f_{2}})^\gamma\big\}^{1/\gamma}$.
Mathematically, it aggregates the difference between $S_{1, f_{1}, f_{2}}$ and $S_{3, f_{1}, f_{2}}$, and the difference between $S_{2, f_{1}, f_{2}}$ and $S_{3, f_{1}, f_{2}}$.
Based on suitable functions $f_1$ and $f_2$,  $\mu_{\gamma,f_{1}, f_{2}}(\x, \y)$ is zero if and only if $\x$ and $\y$ are independent.

For $\gamma=1$, it can be seen that $\mu_{\gamma,f_{1}, f_{2}}(\x, \y)$ reduces to $(S_{1, f_{1}, f_{2}}-S_{3, f_{1}, f_{2}})+(S_{2, f_{1}, f_{2}}-S_{3, f_{1}, f_{2}})$, which is accordant with the existing dependence metrics as mentioned before. However, we notice that under some nonmonotonic and nonlinear relationships, $S_{1, f_{1}, f_{2}}-S_{3, f_{1}, f_{2}}$ and $S_{2, f_{1}, f_{2}}-S_{3, f_{1}, f_{2}}$ are likely to have opposite signs, see simulation results in Table \ref{S123} of Section 4.3. The possible cancellation of positive and neagtive terms would yield small values of
$ \mu_{1,f_{1}, f_{2}}(\x, \y)$, leading to low powers of the tests based on the $ \mu_{1,f_{1}, f_{2}}(\x, \y)$.
In this sense, our newly proposed $\mu_{\gamma,f_{1}, f_{2}}(\x, \y)$ with even and infinity $\gamma$'s can be viewed as significant complements to
$\mu_{1,f_{1}, f_{2}}(\x, \y)$. Theoretically,
we establish the asymptotic null distributions of our proposed test statistics with different $\gamma$ values under the null  hypothesis of independence.
We show that, for  odd $\gamma$ values,  the null limiting distribution is a quadratic function of Gaussian field, which
involves infinitely many nuisance parameters. We suggest to use a  permutation method 
to provide an approximation of the asymptotic null distribution. For even or infinity $\gamma$ values, the null limiting distribution is a weighted  half normal, indicating that our proposed test is asymptotically distribution-free.
Furthermore, we give a comprehensive analysis of how the
value of $ \mu_{\gamma,f_{1}, f_{2}}(\x, \y)$ varies with respect to $\gamma$, under different alternatives characterized by  the signs of $S_{1, f_{1}, f_{2}}-S_{3, f_{1}, f_{2}}$ and $S_{2, f_{1}, f_{2}}-S_{3, f_{1}, f_{2}}$, which forms the basis for power analysis. Since there exists a wide range of dependence scenarios in practice, we further suggest to combine the p-values of test statistics through
Fisher's method \citep{Fisher:1925} to
maintain high power performances.

The rest of the article is organized as follows. In Section 2 we present a general framework to encompass many state-of-the-art dependence metrics in the literature. In Section 3, we propose consistent dependence metrics  by solving a two-sample test  problem.
In Section 4, we develop the tests for independence, and
investigate their
limiting behaviors under the  null  hypothesis of independence and the alternative   of dependence.
In Section 5, we introduce the combined probability tests and a random permutation test procedure.  In Section 6, we  consider  extensions with unknown population distributions. We conduct numerical experiments  in Section 7 to demonstrate the usefulness of our proposed methods, and  provide a brief discussion in Section 8.
The details of  proofs are relegated to an on-line Supplementary Material.

\vspace{-0.5cm}
\section{Preliminaries}
To motivate our proposal, we first revisit the existing dependence metrics
under a unified framework.
For random vectors $\x \in \mR^{d_1}$ and $\y \in \mR^{d_2}$, we
define
$f_{j}$ as the kernel
function of  $m$ arguments in $\calR^{d_{j}}$, for $j\in\{1,2\}$ and integer $m\geq4$.
Using the first $m$ independent copies of $(\x, \y)$, we further define a general metric of dependence as
\beqr\label{gCov}
\mu_{f_{1}, f_{2}}(\x, \y)&\defby& S_{1, f_{1}, f_{2}}+S_{2, f_{1}, f_{2}}-2S_{3, f_{1}, f_{2}},
\eeqr
where $S_{1, f_{1}, f_{2}}$, $S_{2, f_{1}, f_{2}}$ and $S_{3, f_{1}, f_{2}}$ are defined as
\beqrs
S_{1, f_{1}, f_{2}}&\defby &E[\psi_{1,f_{1}, f_{2}}\{(\x_1,\y_1),\ldots,(\x_m,\y_m)\}],\\
S_{2, f_{1}, f_{2}}&\defby &E[\psi_{2,f_{1}, f_{2}}\{(\x_1,\y_1),\ldots,(\x_m,\y_m)\}],\\
S_{3, f_{1}, f_{2}}&\defby &E[\psi_{3,f_{1}, f_{2}}\{(\x_1,\y_1),\ldots,(\x_m,\y_m)\}].
\eeqrs
In the above displays,
\beqrs
\psi_{1, f_{1}, f_{2}}\{(\x_1,\y_1),\ldots,(\x_m,\y_m)\}&\defby& f_{1}(\x_1,\ldots,\x_m)f_{2}(\y_1,\y_2,\y_3,\y_4,\y_5,\ldots,\y_m),
\eeqrs
\beqrs
\psi_{2, f_{1}, f_{2}}\{(\x_1,\y_1),\ldots,(\x_m,\y_m)\}&\defby& f_{1}(\x_1,\ldots,\x_m)f_{2}(\y_3,\y_4,\y_1,\y_2,\y_5,\ldots,\y_m),\\
\psi_{3, f_{1}, f_{2}}\{(\x_1,\y_1),\ldots,(\x_m,\y_m)\}&\defby& f_{1}(\x_1,\ldots,\x_m)f_{2}(\y_1,\y_3,\y_2,\y_4,\y_5,\ldots,\y_m).
\eeqrs
They merely differ in the ordering of $\y_i$s in $f_2$.
It is worth noting that not all  observations of
$\{(\x_i, \y_i), i=1,\ldots,m\}$ must be involved in $f_1$ and $f_2$.

The independence of $\x$ and $\y$ implies $\mu_{f_{1},f_{2}}(\x, \y)=0$
in the family of distributions such that $E(f_{1})$ and $E(f_{2})$ exist and are finite.
This property is referred to as {\it I-consistency} of $\mu_{f_{1}, f_{2}}(\x, \y)$ by \cite{Weihs:Drton:Meinshausen:2018}.
If the dependence of $\x$ and $\y$ implies $\mu_{f_{1},f_{2}}(\x, \y)\neq0$,
then $\mu_{f_{1}, f_{2}}(\x, \y)$ has the {\it D-consistency} property.
It is worth noting that some  choices of $f_{1}$ and $f_{2}$ may make $\mu_{f_{1}, f_{2}}(\x, \y)$
fail to be D-consistent. For instance, in the bivariate case with $d_1=d_2=1$, $\mu_{f_{1},f_{2}}(\x,\y)$ can be simplified to
$\cov^{2}(\x, \y)$ if we set $m=4$, $f_{1}(\x_1,\x_2,\x_3,\x_4)=\x_1\x_2$ and $f_{2}(\y_1,\y_2,\y_3,\y_4)=\y_1\trans\y_2$.
The test
based on such a metric estimator is inconsistent against all
dependent alternatives.
For this reason, we focus on the class of  kernel functions $f_{1}$ and $f_{2}$ so that $\mu_{f_{1},f_{2}}(\x,\y)$ satisfies both I- and D-consistency.

In general, the metric defined in (\ref{gCov}) covers a large number of nonparametric dependence measures.
In the following, we present
five important examples.
Define $\z$ to be $\x$ if $j=1$ and to be $\y$ if $j=2$, and the dimension of $\z$ may differ for $j=1$ and $j=2$.

\textit{Example 2.1} \citep[distance covariance]{Szekely:Rizzo:Bakirov:2007}.
Distance covariance is $\mu_{f_{1},f_{2}}(\x,\y)$ with $m=4$ and $f_{j}^{\dCov}(\z_1,\ldots,\z_4)=\|\z_1-\z_2\|$ for $j\in\{1,2\}$.
Here, $\|\cdot\|$ stands for the Euclidean norm. Thus, we have
$\mu_{f_{1}^{\dCov},f_{2}^{\dCov}}(\x,\y)=E(\|\x_1-\x_2\|\|\y_1-\y_2\|)+E(\|\x_1-\x_2\|\|\y_3-\y_4\|)-2E(\|\x_1-\x_2\|\|\y_1-\y_3\|).$

\textit{Example 2.2} \citep[Gaussian kernel Hilbert-Schmidt independence criterion]{Grettonetal:2008}.
Hilbert-Schmidt independence criterion based on a Gaussian kernel is
$\mu_{f_{1},f_{2}}(\x,\y)$ with $m=4$ and $f_{j}^{\gHSIC}(\z_1,\ldots,\z_4)=\exp\{-\|\z_1-\z_2\|/(2\sigma_{j}^{2})\}$ with  $\sigma_{j}>0$ for $j\in\{1,2\}$.
Thus, we have
{\color{black}$\mu_{f_{1}^{\gHSIC},f_{2}^{\gHSIC}}(\x,\y)=E[\exp\{-\|\x_1-\x_2\|/(2\sigma_{1}^{2})\}\exp\{-\|\y_1-\y_2\|/(2\sigma_{2}^{2})\}]
+E[\exp\{-\|\x_1-\x_2\|/(2\sigma_{1}^{2})\}\exp\{-\|\y_3-\y_4\|/(2\sigma_{2}^{2})\}]-2E[\exp\{-\|\x_1-\x_2\|/(2\sigma_{1}^{2})\}\exp\{-\|\y_1-\y_3\|/(2\sigma_{2}^{2})\}].$}

\textit{Example 2.3} \citep[Heller-Heller-Gorfine coefficient]{Heller:Heller:Gorfine:2013}.
Denote the closed ball
with  center $\z_1$ and  radius $\|\z_1-\z_2\|$ as $B(\z_1, \|\z_1-\z_2\|)$.
Let $I(\cdot)$ be an indicator function.
Heller-Heller-Gorfine coefficient is  $\mu_{f_{1},f_{2}}(\x,\y)$  with $m=6$ and $f_{j}^{\HHG}(\z_1,\ldots,\z_6)=I\{\z_1\in B(\z_5, \|\z_5-\z_6\|)\}I\{\z_2\in B(\z_5, \|\z_5-\z_6\|)\}w_{j}^{\HHG}(\z_1, \z_2)$
for $j\in\{1,2\}$, and $w_{j}^{\HHG}(\z_1, \z_2)\defby[\pr\{\z_3\in B(\z_1, \|\z_1-\z_2\|)\mid \z_1,\z_2\}\pr\{\z_3\not\in B(\z_1, \|\z_1-\z_2\|)\mid \z_1,\z_2\}]^{-1}.$
Thus, we have {\color{black}$\mu_{f_{1}^{\HHG},f_{2}^{\HHG}}(\x,\y)=E[
I\{\x_1\in B(\x_3, \|\x_3-\x_4\|)\}I\{\x_2\in B(\x_3, \|\x_3-\x_4\|)\}I\{\y_1\in B(\y_3, \|\y_3-\y_4\|)\}I\{\y_2\in B(\y_3, \|\y_3-\y_4\|)\}w_{1}^{\HHG}(\x_1, \x_2)w_{2}^{\HHG}(\y_1, \y_2)
]
+
E[
I\{\x_1\in B(\x_5, \|\x_5-\x_6\|)\}I\{\x_2\in B(\x_5, \|\x_5-\x_6\|)\}I\{\y_3\in B(\y_5, \|\y_5-\y_6\|)\}I\{\y_4\in B(\y_5, \|\y_5-\y_6\|)\}w_{1}^{\HHG}(\x_1, \x_2)w_{2}^{\HHG}(\y_3, \y_4)
]$ 
$-
2E[
I\{\x_1\in B(\x_4, \|\x_4-\x_5\|)\}I\{\x_2\in B(\x_4, \|\x_4-\x_5\|)\}I\{\y_1\in B(\y_4, \|\y_4-\y_5\|)\}I\{\y_3\in B(\y_4, \|\y_4-\y_5\|)\}w_{1}^{\HHG}(\x_1, \x_2)w_{2}^{\HHG}(\y_1, \y_3)
]
.$}

\textit{Example 2.4} \citep[projection covariance
]{Zhu:Xu:Li:Zhong:2017}. Let $\arccos(\cdot)$ be the inverse cosine function. Let $\ang(\z_1,\z_2) \defby  \arccos\{(\z_1\trans\z_2)/(\|\z_1\|\|\z_2\|)\}$ stand for the cosine  angle between the random vectors $\z_1$ and $\z_2$.
Under continuity assumption, multivariate Hoeffding's $D$ based on a projection-averaging approach is $\mu_{f_{1},f_{2}}(\x,\y)$ with $m=5$ and
$f_{j}^{\pCov}(\z_1,\ldots,\z_5)=\ang(\z_1-\z_5,\z_2-\z_5)$
for $j\in\{1,2\}$. Thus, we have {\color{black}$\mu_{f_{1}^{\pCov},f_{2}^{\pCov}}(\x,\y)=E\{\ang(\x_1-\x_3,\x_2-\x_3)\ang(\y_1-\y_3,\y_2-\y_3)\}+
E\{\ang(\x_1-\x_5,\x_2-\x_5)\ang(\y_3-\y_5,\y_4-\y_5)\}-2E\{\ang(\x_1-\x_4,\x_2-\x_4)\ang(\y_1-\y_4,\y_3-\y_4)\}.$}

\textit{Example 2.5} \citep[center-outward rank-based distance covariance]{Shi:Drton:Han:2022,Deb:Sen:2023}.
Let $\F_{j,\pm}\in\mR^{d_{j}}$ denote the center-outward distribution function of a continuous random vector $\z$ \citep{Hallin:2017}, then
the center-outward rank-based distance covariance is $\mu_{f_{1},f_{2}}(\x,\y)$  with $m=4$ and $f_{j}^{\RdCov}(\z_1,\ldots,\z_4)$
$=\|\F_{j,\pm}(\z_{1})-\F_{j,\pm}(\z_{2})\|$. Thus, we have
$\mu_{f_{1}^{\RdCov},f_{2}^{\RdCov}}(\x,\y)=E(\|\F_{j,\pm}(\x_{1})-\F_{j,\pm}(\x_{2})\|\|\F_{j,\pm}(\y_{1})-\F_{j,\pm}(\y_{2})\|)+E(\|\F_{j,\pm}(\x_{1})-\F_{j,\pm}(\x_{2})\|\|\F_{j,\pm}(\y_{3})-\F_{j,\pm}(\y_{4})\|)
-2E(\|\F_{j,\pm}(\x_{1})-\F_{j,\pm}(\x_{2})\|\|\F_{j,\pm}(\y_{1})-\F_{j,\pm}(\y_{3})\|).$

\section{Consistent Dependence Metrics through Comparing Two Bivariate Means}
We start from the following theorem.

{\theo{\label{theorem1} {\rm (Equivalence)}}
Assume that kernel functions $f_1$ and $f_2$ enable $\mu_{f_{1}, f_{2}}(\x, \y)$ in  (\ref{gCov}) to be finite and  satisfy both I- and D-consistency.
	Then,  we have $S_{1, f_{1}, f_{2}}$, $S_{2, f_{1}, f_{2}}$ and $S_{3, f_{1}, f_{2}}$ are identical, if and only if $\x$ and $\y$ are independent.
}

Theorem \ref{theorem1} establishes the equivalence between $S_{1, f_{1}, f_{2}}=S_{2, f_{1}, f_{2}}=S_{3, f_{1}, f_{2}}$ and the independence of $\x$ and $\y$, based on suitable kernel functions.
Theorem \ref{theorem1}  inspires us to transform the independence testing problem into checking the equality of two bivariate means, one is $(S_{1, f_{1}, f_{2}}, S_{3, f_{1}, f_{2}})$ and the other is $(S_{3, f_{1}, f_{2}}, S_{2, f_{1}, f_{2}})$.
Therefore, instead of testing the independence of $\x$ and $\y$ directly, we can  test an equivalent
null hypothesis
\beqrs
\hspace{0.5cm} H_0': (S_{1, f_{1}, f_{2}}, S_{3, f_{1}, f_{2}})=(S_{3, f_{1}, f_{2}}, S_{2, f_{1}, f_{2}}), \; \text{versus}\; H_1':(S_{1, f_{1}, f_{2}}, S_{3, f_{1}, f_{2}})\neq(S_{3, f_{1}, f_{2}}, S_{2, f_{1}, f_{2}}).
\eeqrs
\indent We develop a new class of dependence metrics  through solving a two-sample mean testing problem, which is defined as
\beqrs
\mu_{\gamma,f_{1}, f_{2}}(\x, \y)&\defby&\Big\{(S_{1, f_{1}, f_{2}}-S_{3, f_{1}, f_{2}})^\gamma+(S_{2, f_{1}, f_{2}}-S_{3, f_{1}, f_{2}})^\gamma\Big\}^{1/\gamma},
\eeqrs
where the power index {\color{black}$\gamma\in[1, \infty]$ is an integer.} It can be seen that $\mu_{\gamma,f_{1}, f_{2}}(\x, \y)$ measures
the difference between $S_{1, f_{1}, f_{2}}$ and $S_{3, f_{1}, f_{2}}$, as well as the difference between $S_{2, f_{1}, f_{2}}$ and $S_{3, f_{1}, f_{2}}$.
We provide three specific scenarios for $\mu_{\gamma,f_{1}, f_{2}}(\x, \y)$ with different $\gamma$ values.

(i) If $\gamma=1$, we obtain an $\ell_1$-norm-based dependence metric, which is equivalent to $\mu_{f_{1}, f_{2}}(\x, \y)$ defined in (\ref{gCov}).
This implies $\mu_{1,f_{1}, f_{2}}(\x, \y)$ encompasses many popular dependence measures, such as
distance covariance \citep{Szekely:Rizzo:Bakirov:2007}, Heller-Heller-Gorfine coefficient \citep{Heller:Heller:Gorfine:2013}, projection covariance
 \citep{Zhu:Xu:Li:Zhong:2017} and among others.

(ii) If $\gamma=2$, we obtain an $\ell_2$-norm-based dependence metric,
which
can be expressed as
$\mu_{2,f_{1}, f_{2}}(\x, \y)=\{0.5\mu_{f_{1}, f_{2}}(\x, \y)^{2}+0.5(S_{1, f_{1}, f_{2}}-S_{2, f_{1}, f_{2}})^2\}^{1/2}.$

(iii) If $\gamma=\infty$, we obtain a maximum-type measure of dependence metric, which
can be expressed as $\mu_{\infty,f_{1}, f_{2}}(\x, \y)=\max(S_{1, f_{1}, f_{2}}-S_{3, f_{1}, f_{2}}, S_{2, f_{1}, f_{2}}-S_{3, f_{1}, f_{2}}).$
In fact, for any even $\gamma\rightarrow\infty$, we have $\mu_{\gamma,f_{1}, f_{2}}(\x, \y)\rightarrow
	\max(|S_{1, f_{1}, f_{2}}-S_{3, f_{1}, f_{2}}|, |S_{2, f_{1}, f_{2}}-S_{3, f_{1}, f_{2}}|)$.
	Since $(S_{1, f_{1}, f_{2}}-S_{3, f_{1}, f_{2}})+(S_{2, f_{1}, f_{2}}-S_{3, f_{1}, f_{2}})$ is  nonnegative,
	it always holds that
	$\max(|S_{1, f_{1}, f_{2}}-S_{3, f_{1}, f_{2}}|,  |S_{2, f_{1}, f_{2}}-S_{3, f_{1}, f_{2}}|)=\max(S_{1, f_{1}, f_{2}}-S_{3, f_{1}, f_{2}}, S_{2, f_{1}, f_{2}}-S_{3, f_{1}, f_{2}})$.

The following theorem states the consistent properties of our proposed metric.

{\theo{\label{theorem2}{\rm (Consistency)}}
	Assume that kernel functions $f_1$ and $f_2$ enable $\mu_{f_{1}, f_{2}}(\x, \y)$ in  (\ref{gCov}) to be finite, non-negative and satisfy both I- and D-consistency.
	Then, for any integer {\color{black}$\gamma\in[2, \infty]$, we have $ \mu_{\gamma,f_{1}, f_{2}}(\x, \y) \geq 0$, and
	$\mu_{\gamma,f_{1}, f_{2}}(\x, \y)=0$  if and only if $\x$ and $\y$ are independent.}
}

Theorem \ref{theorem2} ensures that $\mu_{\gamma,f_{1}, f_{2}}(\x, \y)$ is able to capture any nonlinear dependence, which lays the foundation for independence testing.
Intuitively, as the value of test statistics increases, there is a stronger evidence to reject the null hypothesis.
For this reason, we first compare the behaviors of  $\mu_{\gamma,f_{1}, f_{2}}(\x, \y)$ with different values of $\gamma$,
at the population level.

{\theo{\label{theorem3} {\rm (Ordering by the magnitude)}}
	Assume that kernel functions $f_1$ and $f_2$ enable $\mu_{f_{1}, f_{2}}(\x, \y)$ in  (\ref{gCov}) to be finite, non-negative and satisfy both I- and D-consistency.
	When $\x$ and $\y$ are dependent,  we make the following conclusions.
	\begin{itemize}
		\item[(i)] If $\gamma$  is even and
		$2\leq\gamma<\infty$, $(S_{1, f_{1}, f_{2}}-S_{3, f_{1}, f_{2}})(S_{2, f_{1}, f_{2}}-S_{3, f_{1}, f_{2}})\neq0$ and either $S_{1, f_{1}, f_{2}}-S_{3, f_{1}, f_{2}}$ or $S_{2, f_{1}, f_{2}}-S_{3, f_{1}, f_{2}}$ is negative, we have
		$$\mu_{\gamma,f_{1}, f_{2}}(\x, \y)>\mu_{\infty,f_{1}, f_{2}}(\x, \y)> \mu_{1, f_{1}, f_{2}}(\x, \y)>0,$$
		and $\mu_{\gamma,f_{1}, f_{2}}(\x, \y)$ is a monotonically decreasing function of $\gamma$;
		\item[(ii)]  If $\gamma$  is odd and
		$2\leq\gamma<\infty$, $(S_{1, f_{1}, f_{2}}-S_{3, f_{1}, f_{2}})(S_{2, f_{1}, f_{2}}-S_{3, f_{1}, f_{2}})\neq0$ and either $S_{1, f_{1}, f_{2}}-S_{3, f_{1}, f_{2}}$ or $S_{2, f_{1}, f_{2}}-S_{3, f_{1}, f_{2}}$ is negative, we have
		$$\mu_{\infty,f_{1}, f_{2}}(\x, \y)>\mu_{\gamma,f_{1}, f_{2}}(\x, \y)> \mu_{1, f_{1}, f_{2}}(\x, \y)>0,$$
		and $\mu_{\gamma,f_{1}, f_{2}}(\x, \y)$ is a monotonically increasing function of $\gamma$;
		\item[(iii)] If both $S_{1, f_{1}, f_{2}}-S_{3, f_{1}, f_{2}}$ and $S_{2, f_{1}, f_{2}}-S_{3, f_{1}, f_{2}}$ are positive, we have
		$$\mu_{1, f_{1}, f_{2}}(\x, \y)>\mu_{\gamma,f_{1}, f_{2}}(\x, \y)> \mu_{\infty,f_{1}, f_{2}}(\x, \y)>0,$$
		for
		$2\leq\gamma<\infty$, and $\mu_{\gamma,f_{1}, f_{2}}(\x, \y)$  is a monotonically decreasing function of $\gamma$;
		\item[(iv)] If either $S_{1, f_{1}, f_{2}}-S_{3, f_{1}, f_{2}}$ or $S_{2, f_{1}, f_{2}}-S_{3, f_{1}, f_{2}}$ is zero, we have
		$$\mu_{1, f_{1}, f_{2}}(\x, \y)=\mu_{\gamma,f_{1}, f_{2}}(\x, \y)= \mu_{\infty,f_{1}, f_{2}}(\x, \y)>0,$$
		for
		$2\leq\gamma<\infty$.
	\end{itemize}
}

The assertions (i), (ii) and (iv) in Theorem \ref{theorem3} illustrate $\mu_{\gamma,f_{1}, f_{2}}$ for $\gamma\geq 2$
is always not smaller than
$\mu_{1, f_{1}, f_{2}}$ in magnitude,  as long as either $S_{1, f_{1}, f_{2}}-S_{3, f_{1}, f_{2}}$ or $S_{2, f_{1}, f_{2}}-S_{3, f_{1}, f_{2}}$ takes negative values.
The intuition behind this is the effects of $S_{1, f_{1}, f_{2}}-S_{3, f_{1}, f_{2}}$ or $S_{2, f_{1}, f_{2}}-S_{3, f_{1}, f_{2}}$
offset each other, leading to a small magnitude of $\mu_{1, f_{1}, f_{2}}$.
The assertion (iii) in Theorem \ref{theorem3} implies that
the converse is trivially true if both $S_{1, f_{1}, f_{2}}-S_{3, f_{1}, f_{2}}$ and $S_{2, f_{1}, f_{2}}-S_{3, f_{1}, f_{2}}$
are positive.
In this sense, the new class of proposed metrics not only inherits the advantages of existing dependence metrics $\mu_{1, f_{1}, f_{2}}$, but also
serves as significant complements under the situations where $S_{1, f_{1}, f_{2}}-S_{3, f_{1}, f_{2}}$ and $S_{2, f_{1}, f_{2}}-S_{3, f_{1}, f_{2}}$ have opposite signs.



\vspace{-0.5cm}
\section{The Proposed   Tests for Independence}
\subsection{The test statistics}
Suppose that $\{(\x_i, \y_i), i=1,\ldots,n\}$ is a random sample drawn from
the population $(\x,\y)$.  
For $l\in\{1,2,3\}$ and $n\geq m$,  an unbiased U-statistic of $S_{l, f_{1}, f_{2}}$ is
\beqrs
\wh S_{l, f_{1}, f_{2}}\defby(n)_m^{-1}\sum\limits_{(i_{1},\ldots,i_{m})}^{n}\psi_{l, f_{1}, f_{2}}\{(\x_{i_1},\y_{i_1}),\ldots,(\x_{i_m},\y_{i_m})\},
\eeqrs
where $(n)_m=n(n-1)\cdots(n-m+1)$ and the summation is taken over  the  indexes that are different from each other.
By replacing $S_{l, f_{1}, f_{2}}$ with $\wh S_{l, f_{1}, f_{2}}$ for $ l\in\{1,2,3\}$,
the empirical versions of $\mu_{\gamma,f_{1}, f_{2}}$ for integer $ \gamma\in[1, \infty]$ is
\beqrs
\wh \mu_{\gamma,f_{1}, f_{2}}(\x, \y)&\defby&\Big\{(\wh S_{1, f_{1}, f_{2}}-\wh S_{3, f_{1}, f_{2}})^\gamma+(\wh S_{2, f_{1}, f_{2}}-\wh S_{3, f_{1}, f_{2}})^\gamma\Big\}^{1/\gamma}, \quad\gamma\in[1, \infty),\\
\wh \mu_{\infty,f_{1}, f_{2}}(\x, \y)&\defby&\max(\wh S_{1, f_{1}, f_{2}}-\wh S_{3, f_{1}, f_{2}}, \wh S_{2, f_{1}, f_{2}}-\wh S_{3, f_{1}, f_{2}}),\quad \gamma=\infty.
\eeqrs
Then we utilize $w_{n, \gamma}\,  \wh \mu_{\gamma,f_{1}, f_{2}}(\x, \y)$ as the
 test statistic for independence, where $w_{n, \gamma}$ is the rate of convergence, defined as
\[
w_{n, \gamma}=
\begin{cases}
	n^{(\gamma+1)/(2\gamma)}, & \text{if $\gamma$ is odd},\\
	n^{1/2}, & \text{if $\gamma$ is  even or $\gamma=\infty$}.
\end{cases}
\]
 To investigate the asymptotic properties of the test statistic, we define
 $\wt \psi_{l, f_{1}, f_{2}}$ for $l=1, 2, 3$ to be the symmetric kernel
\beqrs
\wt \psi_{l, f_{1}, f_{2}}\{(\x_{1},\y_{1}),\ldots,(\x_{m},\y_{m})\}\defby(m)_m^{-1}\sum\limits_{(i_{1},\ldots,i_{m})}^{m}\psi_{l, f_{1}, f_{2}}\{(\x_{i_1},\y_{i_1}),\ldots,(\x_{i_m},\y_{i_m})\}.
\eeqrs
Define the symmetric kernel that is induced by $\wt \psi_{1, f_{1}, f_{2}}$, $\wt \psi_{2, f_{1}, f_{2}}$ and $\wt \psi_{3, f_{1}, f_{2}}$, as
\beqrs
&&h_{f_{1}, f_{2}}\{(\x_{1},\y_{1}),\ldots,(\x_{m},\y_{m})\}\defby\wt \psi_{1, f_{1}, f_{2}}\{(\x_{1},\y_{1}),
\ldots,(\x_{m},\y_{m})\}\\
&&+\wt \psi_{2, f_{1}, f_{2}}\{(\x_{1},\y_{1}),
\ldots,(\x_{m},\y_{m})\}
-2\wt \psi_{3, f_{1}, f_{2}}\{(\x_{1},\y_{1}),
\ldots,(\x_{m},\y_{m})\}.
\eeqrs
For $c\in\{1,\ldots,m\}$ and $l\in\{1, 2, 3\}$, let
\beqrs
h_{c, f_{1}, f_{2}}\{(\x_{1},\y_{1}),\ldots,(\x_{c},\y_{c})\}&\defby&E[h_{f_{1}, f_{2}}\{(\x_{1},\y_{1}),\ldots,(\x_{m},\y_{m})\}\mid \x_1, \ldots,\x_c],\textrm{ and }\\
\wt \psi_{c, l,  f_{1}, f_{2}}\{(\x_{1},\y_{1}),\ldots,(\x_{c},\y_{c})\}&\defby&E[\wt \psi_{l, f_{1}, f_{2}}\{(\x_{1},\y_{1}),\ldots,(\x_{m},\y_{m})\}\mid \x_1, \ldots,\x_c],
\eeqrs
be the respective projections of $h_{f_{1}, f_{2}}$ and $\wt \psi_{l, f_{1}, f_{2}}$.

{\color{black} We focus on a finite candidate set $\Gamma=\{\gamma_{1},\ldots,\gamma_{L}\}$, where $L<\infty$ is a positive integer.
	The following two subsections display the asymptotic  joint distributions of $\{\wh \mu_{\gamma,f_{1}, f_{2}}(\x, \y)\}_{\gamma\in\Gamma}$ under the null and alternative hypotheses.}

\subsection{Joint convergence of $\{\wh \mu_{\gamma,f_{1}, f_{2}}(\x, \y)\}_{\gamma\in\Gamma}$ under independence}
To facilitate the theoretical analysis, we make the following assumptions.

\noindent
\textit{Assumption 1}. Assume that the kernels $\wt \psi_{1, f_{1}, f_{2}}$, $\wt \psi_{2, f_{1}, f_{2}}$ and $\wt \psi_{3, f_{1}, f_{2}}$, induced by the functions $f_1$ and $f_2$, satisfy the following three
properties:
\begin{itemize}
	\item[(i)]  $\wt \psi_{l, f_{1}, f_{2}}\{(\x_{1},\y_{1}),\ldots,(\x_{m},\y_{m})\}$ for $ l\in\{1,2,3\}$ has finite second moments.
	\item[(ii)] $h_{f_{1}, f_{2}}\{(\x_{1},\y_{1}),\ldots,(\x_{m},\y_{m})\}$ has zero-mean and degenerates when $\x$ and $\y$ are independent, that is,
	$h_{1,f_{1}, f_{2}}\{(\x_{1},\y_{1})\}=0$  under independence.
	\item[(iii)] {\color{black}$h_{2,f_{1}, f_{2}}\{(\x_{1},\y_{1}),(\x_{2},\y_{2})\}$
admits the expansion}
	\beqrs
	h_{2,f_{1}, f_{2}}\{(\x_{1},\y_{1}),(\x_{2},\y_{2})\}=\sum\limits_{k=1}^{\infty}\lambda_{k,f_{1}, f_{2}}\phi_{k,f_{1}, f_{2}}(\x_{1},\y_{1})\phi_{k,f_{1}, f_{2}}(\x_{2},\y_{2}),
	\eeqrs
	under independence,
	where $\{\lambda_{k,f_{1}, f_{2}}\}_{k\geq1}$ and $\{\phi_{k,f_{1}, f_{2}}(\cdot,\cdot)\}_{k\geq1}$ are the eigenvalues and orthonormal eigenfunctions of the integral
	equation
	$$E[h_{2,f_{1}, f_{2}}\{(\x_{1},\y_{1}),(\x_{2},\y_{2})\phi_{k,f_{1}, f_{2}}(\x_{2},\y_{2})\}\mid \x_{1},\y_{1}]=\lambda_{k,f_{1}, f_{2}}\phi_{k,f_{1}, f_{2}}(\x_{1},\y_{1}).$$
\end{itemize}
Assumption 1 is rather general. If both $f_1$ and $f_2$ are  square integrable, then this assumption can be satisfied by selecting $f_1$ and $f_2$ from Examples 2.1-2.5.


Define the joint Gaussian random variables $\{\calG_{1,k,f_{1}, f_{2}}\}_{k\geq1}$, and $\calG_{2,f_{1}, f_{2}}$,
where $\calG_{1,k,f_{1}, f_{2}}$ is the $k$-th independent and identically distributed standard normal $N(0,1)$,
and a single variable $\calG_{2,f_{1}, f_{2}}\sim N(0, E[\wt \psi_{1,1, f_{1}, f_{2}}\{(\x_{1},\y_{1})\}-\wt \psi_{1,3, f_{1}, f_{2}}\{(\x_{1},\y_{1})\}]^{2})$.
The covariance
between  $\calG_{1,k,f_{1}, f_{2}}$ and $\calG_{2,f_{1}, f_{2}}$ is defined as
$\cov(\calG_{1,k,f_{1}, f_{2}}, \calG_{2,f_{1}, f_{2}})=E\{\phi_{k,f_{1}, f_{2}}(\x_{1},\y_{1})
[\wt \psi_{1,1, f_{1}, f_{2}}\{(\x_{1},\y_{1})\}$ $-\wt \psi_{1,3, f_{1}, f_{2}}\{(\x_{1},\y_{1})\}]\}.$

%
%
%
%
%

%

{\theo{\label{theorem4}  {\rm (Joint limiting null distribution)}}
	Suppose that Assumption 1 is
	fulfilled.
	When $\x$ and $\y$ are independent,
the test statistics	$\{w_{n, \gamma}\, \wh \mu_{\gamma,f_{1}, f_{2}}(\x, \y)\}_{\gamma\in\Gamma}$ converge in distribution to
	$\{W_{0, \gamma,f_{1}, f_{2}}\}_{\gamma\in\Gamma}$,
	where
	\[
	W_{0, \gamma,f_{1}, f_{2}}\defby
	\begin{cases}
		m\{(m-1)\gamma/2\}^{1/\gamma}\Big\{\sum\limits_{k=1}^{\infty}\lambda_{k,f_{1}, f_{2}}(\calG_{1,k,f_{1}, f_{2}}^{2}-1)\Big\}^{1/\gamma}\calG_{2,f_{1}, f_{2}}^{1-1/\gamma}, & \text{if $\gamma$ is odd},\\
		m2^{1/\gamma}| \calG_{2,f_{1}, f_{2}}|, & \text{if $\gamma$ is  even},\\
		m| \calG_{2,f_{1}, f_{2}}|, & \text{if $\gamma=\infty$}.
	\end{cases}
	\]
}
Theorem \ref{theorem4} delivers several pieces of information. First, the test statistics $w_{n, \gamma}\, \wh \mu_{\gamma,f_{1}, f_{2}}(\x, \y),$
 $\gamma\in\Gamma$ are not asymptotically independent across different values of $\gamma$'s.
Second, for $\gamma=1$, the original test statistic $n \,\wh \mu_{1,f_{1}, f_{2}}(\x, \y)$ is asymptotically distributed as
$\{m(m-1)/2\}\sum_{k=1}^{\infty}\lambda_{k,f_{1}, f_{2}}$ $(\calG_{1,k,f_{1}, f_{2}}^{2}-1)$, which is consistent with the conclusions drawn
from the literature for Examples 2.1-2.5.
Thrid, for an odd $\gamma$, the limiting null distirbution $W_{0, \gamma,f_{1}, f_{2}}$  involves infinitely many unknown
parameters $\{\lambda_{k,f_{1}, f_{2}}\}_{k\geq1}$, making it intractable. In practice, a permutation procedure can be applied to obtain the p-values.
Its  consistency will be established in Theorem \ref{theorem6} of Section \ref{sec:per}.
Fourth, for an even or infinity $\gamma$, the test statistic $w_{n, \gamma}\, \wh \mu_{\gamma,f_{1}, f_{2}}(\x, \y)$  is asymptotically distributed as a weighted half normal, indicating that
the test based on $w_{n, \gamma}\, \wh \mu_{\gamma,f_{1}, f_{2}}(\x, \y)$ is asymptotically distribution-free. To approximate the null distribution, we need to provide a consistent estimate for the variance of $\calG_{2,f_{1}, f_{2}}$.
We employ a jackknife estimate, given by
\beqrs
\wh\sigma_{0,f_{1}, f_{2}}^{2}&\defby&(n-1)(n-m)^{-2}\sum\limits_{i=1}^{n}\Bigg\{(n-1)_{m-1}^{-1}\sum_{\substack{
		(i_{1},\ldots,i_{m-1})\\
		i\not\in\{i_{1},\ldots,i_{m-1}\}
}}^{n}\Bigg[\wt \psi_{1,1, f_{1}, f_{2}}\{(\x_{i},\y_{i}),(\x_{i_1},\y_{i_1})\ldots,\\
&&(\x_{i_{m-1}},\y_{i_{m-1}})\}
-\wt \psi_{1,3, f_{1}, f_{2}}\{(\x_{i},\y_{i}),(\x_{i_1},\y_{i_1})\ldots,(\x_{i_{m-1}},\y_{i_{m-1}})\}\Bigg]\Bigg\}^2.
\eeqrs
The  convergence of U-statistics \citep{Korolyuk:Borovskich:2013} guarantees that
$\wh\sigma_{0,f_{1}, f_{2}}^{2}$ is a consistent estimate. Therefore, by applying Theorem \ref{theorem4} and the Slutsky theorem, we further obtain
$n^{1/2}\wh \mu_{\gamma,f_{1}, f_{2}}(\x, \y)/\{2^{1/\gamma}m\wh\sigma_{0,f_{1}, f_{2}}\}\sim | N(0 ,1)|$
   under the independence, for even or infinity $\gamma$.
   This provides a computationally efficient way to compute p-values.


\vspace{-0.5cm}
\subsection{Joint convergence of $\{\wh \mu_{\gamma,f_{1}, f_{2}}(\x, \y)\}_{\gamma\in\Gamma}$ under dependence}
In this section, we turn to analyze the asymptotic power and assume the following condition.
\noindent
\textit{Assumption 2}. Assume that the kernels $\wt \psi_{1, f_{1}, f_{2}}$, $\wt \psi_{2, f_{1}, f_{2}}$ and $\wt \psi_{3, f_{1}, f_{2}}$, induced by the functions $f_1$ and $f_2$, satisfy the following two
properties:
\begin{itemize}
	\item[(i)]  $\wt \psi_{l, f_{1}, f_{2}}\{(\x_{1},\y_{1}),\ldots,(\x_{m},\y_{m})\}$  have finite second moments.
	\item[(ii)] $h_{f_{1}, f_{2}}\{(\x_{1},\y_{1}),\ldots,(\x_{m},\y_{m})\}$ has positive mean and does not degenerate when $\x$ and $\y$ are not independent, that is,
	$\var\{h_{1,f_{1}, f_{2}}(\x_{1},\y_{1})\}>0$  under dependence.
\end{itemize}

Assumption 2 is also satisfied for square integrable functions, such as $f_1$ and $f_2$ in Examples 2.1-2.5.
We further define the joint Gaussian random variables $\calG_{3,f_{1}, f_{2}}$ and $\calG_{4,f_{1}, f_{2}}$,
where
$\calG_{3,f_{1}, f_{2}}\sim N(0, \var\{\wt \psi_{1,1, f_{1}, f_{2}}(\x_{1},\y_{1})-\wt \psi_{1,3, f_{1}, f_{2}}(\x_{1},\y_{1})\})$,
$\calG_{4,f_{1}, f_{2}}\sim N(0, \var\{\wt \psi_{1,2, f_{1}, f_{2}}(\x_{1},$ $\y_{1})-\wt \psi_{1,3, f_{1}, f_{2}}(\x_{1},\y_{1})\})$
and their covariance is given by
${\cov(\calG_{3,f_{1}, f_{2}}},$
$ \calG_{4,f_{1}, f_{2}})=\cov\{\wt \psi_{1,1, f_{1}, f_{2}}($
$\x_{1},\y_{1})-\wt \psi_{1,3, f_{1}, f_{2}}(\x_{1},\y_{1}),\ 
\wt \psi_{1,2, f_{1}, f_{2}}(\x_{1},\y_{1})-\wt \psi_{1,3, f_{1}, f_{2}}(\x_{1},\y_{1})$\}.

In  Theorem \ref{theorem5}, we give an explicit description for  the asymptotic joint distribution of  $\wh \mu_{\gamma,f_{1}, f_{2}}(\x, \y)$ for  $1\leq\gamma<\infty$, and $\wh \mu_{\infty,f_{1}, f_{2}}(\x, \y)$ under general dependencies.

{\theo{\label{theorem5}} {\rm (Joint limiting alternative distribution)}
	Suppose Assumption 2 is
	fulfilled.
	When $\x$ and $\y$ are dependent,
	 the test statistics $\{n^{1/2}\, \wh \mu_{\gamma,f_{1}, f_{2}}(\x, \y)-n^{1/2}\,\mu_{\gamma,f_{1}, f_{2}}(\x, \y)\}_{\gamma\in\Gamma}$ converge in distribution to
	$\{W_{1, \gamma,f_{1}, f_{2}}\}_{\gamma\in\Gamma}$,
	where
	\[
	W_{1, \gamma,f_{1}, f_{2}}\defby
	\begin{cases}
		&m\mu_{\gamma,f_{1}, f_{2}}^{1-\gamma}\Big\{(S_{1, f_{1}, f_{2}}-S_{3, f_{1}, f_{2}})^{\gamma-1}\calG_{3,f_{1}, f_{2}}+(S_{2, f_{1}, f_{2}}-S_{3, f_{1}, f_{2}})^{\gamma-1}\calG_{4,f_{1}, f_{2}}\Big\}, \\
		&\hspace{9cm} \text{if $\gamma$ is odd or even},\\
		&m\calG_{3,f_{1}, f_{2}}I(S_{1, f_{1}, f_{2}}>S_{2, f_{1}, f_{2}})+m\calG_{4,f_{1}, f_{2}}I(S_{1, f_{1}, f_{2}}<S_{2, f_{1}, f_{2}})\\
		&+m\max(\calG_{3,f_{1}, f_{2}}, \calG_{4,f_{1}, f_{2}}) I(S_{1, f_{1}, f_{2}}=S_{2, f_{1}, f_{2}}\neq S_{3, f_{1}, f_{2}}), \quad \text{if $\gamma=\infty$}.
	\end{cases}
	\]
}
   Theorem \ref{theorem4}  and Corollary \ref{corollary1} indicate that $w_{n, \gamma}\ \wh \mu_{\gamma,f_{1}, f_{2}}(\x, \y)$ can
serve as a  statistic  to test for the independence, which is consistent against general alternatives.	

{\coll{\label{corollary1}} {\rm (Consistency)}
Suppose all assumptions in Theorem \ref{theorem2} and Theorem \ref{theorem5} hold.
When $\x$ and $\y$ are dependent, for any $\gamma\in\Gamma$,  it always holds that
\begin{itemize}
	\item[(i)] $\wh \mu_{\gamma,f_{1}, f_{2}}(\x, \y)\rightarrow\mu_{\gamma,f_{1}, f_{2}}(\x, \y)>0$ in probability;
	\item[(ii)] $w_{n, \gamma}\ \wh \mu_{\gamma,f_{1}, f_{2}}(\x, \y)\rightarrow\infty$ in probability.
\end{itemize}
}
Based on Theorems  \ref{theorem3}--\ref{theorem4} and Corollary \ref{corollary1},
we provide power analysis for the test statistics $w_{n, \gamma}\, \wh \mu_{\gamma,f_{1}, f_{2}}(\x, \y)$ with  different values of $\gamma$. Denote the observations $\calD_n \defby \{(\x_i,\y_i), i=1,\cdots,n\}$. Let  $\wh p_{\gamma,f_{1}, f_{2}}(\x, \y)$ be the p-value of the test based on $w_{n, \gamma}\, \wh \mu_{\gamma,f_{1}, f_{2}}(\x, \y)$, given by
\beqrs
\wh p_{\gamma,f_{1}, f_{2}}(\x, \y)\defby
\begin{cases}
	\pr\Big(\{(m-1)\gamma/2\}^{1/\gamma}\Big\{\sum\limits_{k=1}^{\infty}\lambda_{k,f_{1}, f_{2}}(\calG_{1,k,f_{1}, f_{2}}^{2}-1)\Big\}^{1/\gamma}\calG_{2,f_{1}, f_{2}}^{1-1/\gamma} &\\
>w_{n, \gamma}\, \wh \mu_{\gamma,f_{1}, f_{2}}(\x, \y)/m \mid \calD_n\Big),
	& \text{if $\gamma$ is odd},\\
	\pr\{| \calG_{2,f_{1}, f_{2}}|>w_{n, \gamma}\, \wh \mu_{\gamma,f_{1}, f_{2}}(\x, \y)/(m2^{1/\gamma}) \mid \calD_n\}, & \text{if $\gamma$ is  even},\\
	\pr\{| \calG_{2,f_{1}, f_{2}}|>w_{n, \gamma}\, \wh \mu_{\gamma,f_{1}, f_{2}}(\x, \y)/m \mid \calD_n\} & \text{if $\gamma=\infty$}.
\end{cases}
\eeqrs

Intuitively, the smaller the p-value, the more compelling the evidence to reject the null hypothesis.
We first consider the scenario where $\gamma$ goes to infinity. For  an odd $\gamma$, as $\gamma\rightarrow\infty$,
\beqrs
\{(m-1)\gamma/2\}^{1/\gamma}\Big\{\sum\limits_{k=1}^{\infty}\lambda_{k,f_{1}, f_{2}}(\calG_{1,k,f_{1}, f_{2}}^{2}-1)\Big\}^{1/\gamma}\calG_{2,f_{1}, f_{2}}^{1-1/\gamma}\rightarrow|\calG_{2,f_{1}, f_{2}}|,
\eeqrs
in distribution. Accordingly,  the p-value for odd $\gamma\rightarrow\infty$ can be further simplified to
$\pr\{| \calG_{2,f_{1}, f_{2}}|>w_{n, \gamma}\, \wh \mu_{\gamma,f_{1}, f_{2}}(\x, \y)/m \mid \calD_n\}.$
Since $w_{n, \gamma}/m$ and $w_{n, \gamma}/(m2^{1/\gamma}) $ are almost the same as $\gamma\rightarrow\infty$,
comparing the values of p-value is equivalently transformed into comparing the magnitudes of $ \wh \mu_{\gamma,f_{1}, f_{2}}(\x, \y)$.
From  Theorem  \ref{theorem3} and Corollary \ref{corollary1},  no matter $\gamma$ is even or odd,
$\pr\{\wh \mu_{\gamma,f_{1}, f_{2}}(\x, \y)>t\}$ is a monotonic function of $\gamma$ for any fixed $t\in\mR^1$.
Therefore, there must exist some sufficiently large $\gamma_0$, such that  $\pr\{\wh \mu_{\gamma,f_{1}, f_{2}}(\x, \y)>t\}$ reaches its maximum at either $\gamma=\gamma_0$ or $\gamma=\infty$.
This implies it is reasonable to focus on a finite candidate set $\Gamma=\{\gamma_{1},\ldots,\gamma_{L}\}$, rather than an infinite one.

For any finite $\gamma$, the statistic $w_{n, \gamma}\, \wh \mu_{\gamma,f_{1}, f_{2}}(\x, \y)$ is of order $O(1)$   under the independence, as shown in Theorem \ref{theorem4}.
We need to consider the simultaneous variation of $w_{n, \gamma}$ and $\wh \mu_{\gamma,f_{1}, f_{2}}(\x, \y)$ with respect to  $\gamma$ under the dependence.
For an odd $\gamma$, $w_{n, \gamma}$ is a monotonically decreasing function of $\gamma$.
Under the altervatives where both $S_{1, f_{1}, f_{2}}-S_{3, f_{1}, f_{2}}$ and $S_{2, f_{1}, f_{2}}-S_{3, f_{1}, f_{2}}$ are positive,
$\pr\{\wh \mu_{\gamma,f_{1}, f_{2}}(\x, \y)>t\}$ is also a monotonically decreasing function of $\gamma$ for any fixed $t\in\mR^1$, which leads to the test with $\gamma=1$ being most powerful.
However, under the altervatives where either $S_{1, f_{1}, f_{2}}-S_{3, f_{1}, f_{2}}$ or $S_{2, f_{1}, f_{2}}-S_{3, f_{1}, f_{2}}$ is negative, $\pr\{\wh \mu_{\gamma,f_{1}, f_{2}}(\x, \y)>t\}$ becomes a monotonically increasing function of $\gamma$ for any fixed $t\in\mR^1$.
The opposite trends of $w_{n, \gamma}$ and $\wh \mu_{\gamma,f_{1}, f_{2}}(\x, \y)$ make the test achieve the best power with some odd $\gamma_0$.
For an even $\gamma$, $w_{n, \gamma}/(2^{1/\gamma})$ is a monotonically increasing function of $\gamma$, while
$\pr\{\wh \mu_{\gamma,f_{1}, f_{2}}(\x, \y)>t\}$ has an opposite trend concerning $\gamma$, under the altervatives where either $S_{1, f_{1}, f_{2}}-S_{3, f_{1}, f_{2}}$ or $S_{2, f_{1}, f_{2}}-S_{3, f_{1}, f_{2}}$ is negative. This implies,
there also exists some $\gamma_0$ that enables the asymptotic power based on $w_{n, \gamma_0}\, \wh \mu_{\gamma_0, f_{1}, f_{2}}(\x, \y)$ to perform the best.

Based on
the aforementioned theoretical analysis,
we recommend to include small  values such as 1 and 2, medium ones such as $3,\ldots,6$, and  extremely large values  such as $\infty$ into $\Gamma$ to balance the asymptotic and non-asymptotic performances.



\begin{figure}[htp!]
	\centerline{
		\begin{tabular}{ccc}
			\psfig{figure=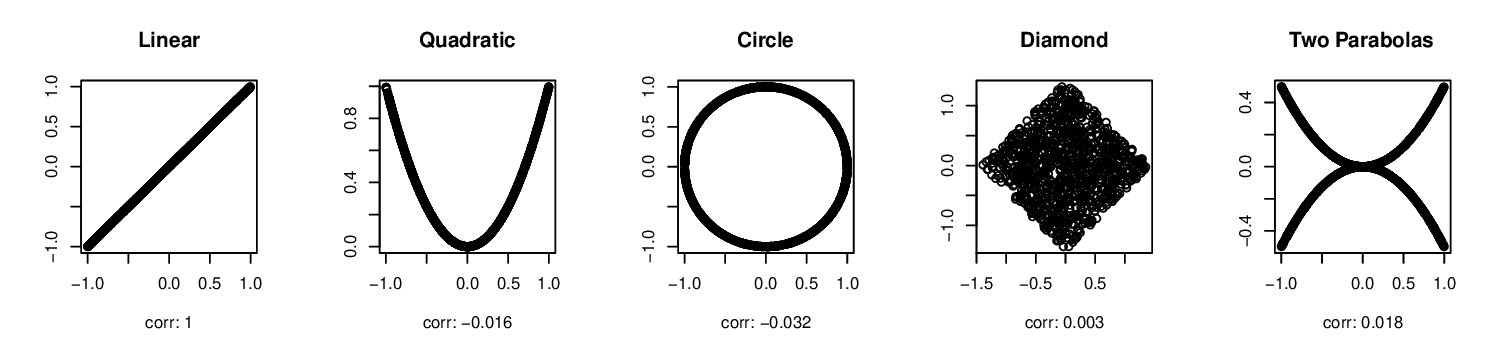,width=6.5in,height=1.5in,angle=0}
	\end{tabular}}
	\caption{\label{fig:rho}  For each panel, a pair of dependent $(X, Y)$ is generated without any noise and visualized based on
		1,000 random sample generations. Pearson correlation is displayed at the bottom of plot.
	}
\end{figure}

We use simulation examples to demonstrate the dependence alternatives described by $S_{1, f_{1}, f_{2}}-S_{3, f_{1}, f_{2}}$ and $S_{2, f_{1}, f_{2}}-S_{3, f_{1}, f_{2}}$,
since the closed forms of $S_{1, f_{1}, f_{2}}-S_{3, f_{1}, f_{2}}$ and $S_{2, f_{1}, f_{2}}-S_{3, f_{1}, f_{2}}$ are not easy to obtain.
We consider the five models employed in Section \ref{sec:sim}.
To visualize the relationships in these models, we generate the paris of $(X,Y)$, $d_1=d_2=1$ without any noise for each configuration and provide a scatter plot in Figure \ref{fig:rho}.
It can be observed that  Pearson correlation equals to one when there exists a linear association, and
almost zero for the nonmonotonic and nonlinear relationships. Furthermore, we calculate the values of $S_{1, f_{1}, f_{2}}-S_{3, f_{1}, f_{2}}$ and $S_{2, f_{1}, f_{2}}-S_{3, f_{1}, f_{2}}$
with the kernel functions $f_1=f_2=f^{\dCov}$, and summarize the results in Table \ref{S123}.
For the linear model (M1), the value of $S_{1, f_{1}, f_{2}}+S_{2, f_{1}, f_{2}}-2S_{3, f_{1}, f_{2}}$
exhibit a significant positive improvement compared to the null hypothesis, which helps the testing methods
maintain high efficiency to detect the linear  relationships.
For the quadratic model (M2), the value of $S_{1, f_{1}, f_{2}}+S_{2, f_{1}, f_{2}}-2S_{3, f_{1}, f_{2}}$ is slightly larger than zero when $d_1=d_2=5$, but decreases rapidly to zero as the dimensions increase.
For other nonmonotonic and nonlinear  models (M3)-(M5), $S_{1, f_{1}, f_{2}}-S_{3, f_{1}, f_{2}}$ and  $S_{2, f_{1}, f_{2}}-S_{3, f_{1}, f_{2}}$ have
opposite signs and their absolute values are
mostly comparable, which making the value of $S_{1, f_{1}, f_{2}}+S_{2, f_{1}, f_{2}}-2S_{3, f_{1}, f_{2}}$ close to zero.
This cancellation effect hinders the power performances of the testing methods based $\wh \mu_{1,f_{1}, f_{2}}(\x, \y)$.

\begin{table}[h!]
	\begin{center}
		\caption{\label{S123} $S_{1}-S_{3}$, $S_{2}-S_{3}$ and their summation under $d_1=d_2=\{5,100,200,400\}$ with normal and $t(3)$ errors.}
				\footnotesize
		\begin{tabular}{lcrrrrrrrrr}
			\toprule
			&& \multicolumn{4}{c}{Normal}&&\multicolumn{4}{c}{$t(3)$} \\\cline{3-6}\cline{8-11}
		Model	&&5&100&200&400&&5&100&200&400\\\hline
		   	\multirow{2}{*}{(M1)}   &$S_{1}-S_{3}$           &0.073 & 0.072 & 0.072 & 0.071 && 0.142 & 0.132 & 0.129 & 0.131 \\
		                                    &$S_{2}-S_{3}$   & -0.012 & -0.012 & -0.012 & -0.012 && -0.023 & -0.022 & -0.021 & -0.023 \\
		                             &$S_{1}+S_{2}-2S_{3}$   &  0.061 & 0.060 & 0.060 & 0.059 && 0.119 & 0.110 & 0.108 & 0.108 \\   \hline
		                             \multirow{2}{*}{(M2)}   &$S_{1}-S_{3}$           & 0.018 & 0.012 & 0.012 & 0.012 && 0.018 & 0.012 & 0.012 & 0.012 \\
		                             &$S_{2}-S_{3}$   & -0.012 & -0.012 & -0.012 & -0.012 && -0.012 & -0.012 & -0.012 & -0.012\\
		                             &$S_{1}+S_{2}-2S_{3}$   & 0.006 & 0.000 & 0.000 & 0.000 && 0.006 & 0.000 & 0.000 & 0.000 \\ \hline
		                             \multirow{2}{*}{(M3)}   &$S_{1}-S_{3}$           & -0.025 & -0.025 & -0.025 & -0.025 && -0.028 & -0.029 & -0.029 & -0.029\\
		                             &$S_{2}-S_{3}$   & 0.025 & 0.025 & 0.026 & 0.025 && 0.028 & 0.029 & 0.030 & 0.030 \\
		                             &$S_{1}+S_{2}-2S_{3}$   & 0.000 & 0.000 & 0.001 & 0.000 && 0.000 & 0.000 & 0.001 & 0.001 \\  \hline
		                             \multirow{2}{*}{(M4)}   &$S_{1}-S_{3}$           & -0.024 & -0.025 & -0.025 & -0.025 && -0.024 & -0.025 & -0.025 & -0.025 \\
		                             &$S_{2}-S_{3}$   & 0.024 & 0.025 & 0.025 & 0.025 && 0.024 & 0.025 & 0.025 & 0.025 \\
		                             &$S_{1}+S_{2}-2S_{3}$   & 0.000 & 0.000 & 0.000 & 0.000 && 0.000 & 0.000 & 0.000 & 0.000 \\  \hline
		
		                             \multirow{2}{*}{(M5)}   &$S_{1}-S_{3}$           & 0.012 & 0.012 & 0.012 & 0.012 && 0.018 & 0.017 & 0.017 & 0.017 \\
		                             &$S_{2}-S_{3}$   & -0.011 & -0.012 & -0.012 & -0.012 && -0.018& -0.017 & -0.017 & -0.017\\
		                             &$S_{1}+S_{2}-2S_{3}$   & 0.001 & 0.000 & 0.000 & 0.000 && 0.000 & 0.000 & 0.000 & 0.000 \\  \hline
		\end{tabular}
	\end{center}
\end{table}

\section{Combined Probability Test and Permutation Test}\label{sec:per}
In practice, since we target at a wide range of dependence alternatives,  we suggest to combine the test statistics $w_{n, \gamma}\, \wh \mu_{\gamma,f_{1}, f_{2}}(\x, \y), \gamma\in\Gamma$, through their respective p-values, to introduce an aggregated test  of the independence.
There are several p-value combination methods available to use. For example, we employ the Fisher's method \citep{Fisher:1925}, then the test statistic is defined as
\beqrs
\wh T_{\text{Fisher}}(\x,\y)\defby\sum\limits_{\gamma\in\Gamma}-2\log \wh p_{\gamma,f_{1}, f_{2}}(\x,\y).
\eeqrs
Fisher's method achieves the asymptotic optimality with respect to Bahadur
relative efficiency \citep{Littell:Folks:1971,Littell:Folks:1973}. If we adopt the minimum combination \citep{Xu:Lin:Wei:Pan:2016,He:Xu:Wu:Pan:2021}, the test statistic is given by
\beqrs
\wh T_{\text{min}}(\x,\y)\defby -\min_{\gamma\in\Gamma}\wh p_{\gamma,f_{1}, f_{2}}(\x,\y).
\eeqrs
The test statistic based on the Cauchy combination \citep{Liu:Xie:2020} is
\beqrs
\wh T_{\text{Cauchy}}(\x,\y)\defby\sum\limits_{\gamma\in\Gamma}\frac12\tan\left[\pi\left\{\frac12-\wh p_{\gamma,f_{1}, f_{2}}(\x,\y)\right\}\right].
\eeqrs
We report the numerical results based on these three types of combination methods in Section \ref{sec:num}, which do not reveal the significant differences. Thus, we can simply
use the Fisher's method in practice.

By Theorem \ref{theorem4},  $w_{n, \gamma}\, \wh \mu_{\gamma,f_{1}, f_{2}}(\x, \y), \gamma\in\Gamma$ are not asymptotically
independent across different values of $\gamma$, and their asymptotic null distributions are not   tractable if $\Gamma$ contains odd numbers.
A permutation procedure is useful to approximate the null
distributions of test statistics and find the corresponding p-values.


\begin{algorithm}[!ht]
	\caption{The implementation of permutation-based  procedure}
	\label{power1}
	\begin{algorithmic} 
		\STATE {\textbf{Step 1}}: Given the sample $\{(\x_i, \y_i), i=1,\ldots,n\}$ and kernel functions $f_1$ and $f_2$, compute the test statistics $\{w_{n, \gamma}\, \wh \mu_{\gamma,f_{1}, f_{2}}(\x, \y)\}_{\gamma \in \Gamma}$.
		\STATE {\textbf{Step 2}}:  For $b=1,\cdots, B$, generate a random permutation of $\{1,2,\ldots,n\}$, defined as $\{j_1,j_2,\ldots,j_n\}$.
		Then the permutation sample $\y^b_i \defby \y_{j_i}$,  $i\in\{1,\ldots,n\}$. Re-estimate the test statistics $\{w_{n, \gamma}\, \wh \mu_{\gamma,f_{1}, f_{2}}(\x, \y^b)\}_{\gamma \in \Gamma}$.
	\STATE {\textbf{Step 3}}: 	We approximate the p-values $\wh p_{\gamma,f_{1}, f_{2}}(\x, \y), \gamma\in\Gamma$ with permutation probabilities
		\beqrs
		\wt p_{\gamma,f_{1}, f_{2}}(\x, \y)=B^{-1}\sum_{b_0=1}^B I\left\{w_{n, \gamma}\, \wh \mu_{\gamma,f_{1}, f_{2}}(\x, \y^{b_0})>w_{n, \gamma}\, \wh \mu_{\gamma,f_{1}, f_{2}}(\x, \y)\mid \calD_n\right\}, \gamma\in\Gamma.
		\eeqrs
		The p-values $\wt p_{\gamma,f_{1}, f_{2}}(\x, \y^{b}), \gamma\in\Gamma$ can be further calculated.
		\STATE {\textbf{Step 4}}: Compute the combined test statistics $\wh T_{\Gamma,f_{1}, f_{2}}(\x, \y)$ and
		$\wh T_{\Gamma,f_{1}, f_{2}}(\x, \y^{b})$ based on the p-values $\wt p_{\gamma,f_{1}, f_{2}}(\x, \y)$ and $\wt p_{\gamma,f_{1}, f_{2}}(\x, \y^{b})$ in Step 3, where $\wh T\in\{\wh T_{\text{Fisher}},\wh T_{\text{min}},\wh T_{\text{Cauchy}}\}$.
		\STATE {\textbf{Step 5}}:
		The corresponding p-values of $\wh T_{ \Gamma, f_{1}, f_{2}}(\x, \y)$ can be approximated with
			\beqrs
		B^{-1}\sum_{b=1}^B I\left\{\wh T_{\Gamma,f_{1}, f_{2}}(\x, \y^{b})>\wh T_{\Gamma,f_{1}, f_{2}}(\x, \y)\mid \calD_n\right\}.
		\eeqrs
		
	\end{algorithmic}
\end{algorithm}

Theorem \ref{theorem6} shows that the joint distribution of $\{w_{n, \gamma}\, \wh \mu_{\gamma,f_{1}, f_{2}}(\x, \y^{b})\}_{\gamma\in\Gamma}$,  given the observations $\calD_n$,
converges to the asymptotic null distribution  of $\{w_{n, \gamma}\,\wh \mu_{\gamma,f_{1}, f_{2}}(\x, \y)\}_{\gamma\in\Gamma}$.
Therefore, the permutation procedure is consistent.
To prove this result, it is necessary to assume the finite $(2+\delta)$ moment condition, which is similar to, but slightly stronger than, those specified in Assumptions 1-2.
Using  a higher-order moment condition  is essential to verify  the uniform integrability for a  class of random variables.

\noindent
\textit{Assumption 3}. For any $1\leq i_1,\ldots,i_m\leq m$, there exists $\delta>0$ such that the kernels
$\wt \psi_{l, f_{1}, f_{2}}\{(\x_{1},\y_{i_1}),\ldots,(\x_{m},\y_{i_m})\},$
$ l\in\{1,2,3\}$, have finite  $(2+\delta)$ moments.

{\theo{\label{theorem6}} {\rm (Limiting distributions of permutated statistics)}
Suppose that Assumptions 1 and 3 are
fulfilled, we have
\begin{itemize}
	\item[(i)] If $\x$ and $\y$ are independent, conditional on the observations,  $\{w_{n, \gamma}\, \wh \mu_{\gamma,f_{1}, f_{2}}(\x, \y^{b})\}_{\gamma\in\Gamma}$ converges weakly to
	$\{W_{0, \gamma,f_{1}, f_{2}}\}_{\gamma\in\Gamma}$, where $\{W_{0, \gamma,f_{1}, f_{2}}\}_{\gamma\in\Gamma }$ are given in Theorem \ref{theorem4};
	\item[(ii)] If $\x$ and $\y$ are dependent, conditional on the observations,  $\{w_{n, \gamma}\, \wh \mu_{\gamma,f_{1}, f_{2}}(\x, \y^{b})\}_{\gamma\in\Gamma}$ converges weakly to
	$\{\wt W_{0, \gamma,f_{1}, f_{2}}\}_{\gamma\in\Gamma}$, which is the asymptotic joint distribution of $\{w_{n, \gamma}$ $\wh \mu_{\gamma,f_{1}, f_{2}}(\x, \wt\y)\}_{\gamma\in\Gamma}$.
	Here, $\wh \mu_{\gamma,f_{1}, f_{2}}(\x, \wt\y)$ is calculated through  independent observations $\{(\x_i,\wt \y_i),i=1,
	\ldots,n\}$, where $\wt \y_i$ is an independent copy of $\y_i$, and more importantly, $\wt \y_i$  is   independent of $\x_i$.
\end{itemize}
}

A direct application of Theorems \ref{theorem4}-\ref{theorem6} yields the following corollary.

{\coll{\label{corollary3}} {\rm (Consistency of permutation)}
Suppose that Assumptions 1-3 are
fulfilled.
\begin{itemize}
	\item[(i)] If $\x$ and $\y$ are independent,  we have the permutation  p-values $\pr\{w_{n, \gamma}\, \wh \mu_{\gamma,f_{1}, f_{2}}(\x, \y^{b})>$
	$w_{n, \gamma}\, \wh \mu_{\gamma,f_{1}, f_{2}}(\x, \y)\mid \calD_n\},\gamma\in\Gamma$ and $\pr\{ \wh T_{\Gamma,f_{1}, f_{2}}(\x, \y^{b})>\wh T_{\Gamma,f_{1}, f_{2}}(\x, \y)\mid \calD_n\}, \wh T\in\{\wh T_{\text{Fisher}},\wh T_{\text{min}},\wh T_{\text{Cauchy}}\}$,  converge weakly to the uniform distribution on $(0, 1)$;
	\item[(ii)] If $\x$ and $\y$ are dependent,   we have the permutation  p-values $\pr\{w_{n, \gamma}\ \wh \mu_{\gamma,f_{1}, f_{2}}(\x, \y^{b})>$
	$w_{n, \gamma}\ \wh \mu_{\gamma,f_{1}, f_{2}}(\x, \y)\mid \calD_n\},\gamma\in\Gamma$ and $\pr\{ \wh T_{\Gamma,f_{1}, f_{2}}(\x, \y^{b})>\wh T_{\Gamma,f_{1}, f_{2}}(\x, \y)\mid \calD_n\}, \wh T\in\{\wh T_{\text{Fisher}},\wh T_{\text{min}},\wh T_{\text{Cauchy}}\}$,  converge  to zero in probability.
\end{itemize}}

The assertion (i) in Corollary \ref{corollary3} implies that  the proposed permutation tests have a correct asymptotic level.
The assertion (ii) in Corollary \ref{corollary3} shows that the tests based on
$w_{n, \gamma}\ \wh \mu_{\gamma,f_{1}, f_{2}}(\x, \y), \gamma\in\Gamma$ and $\wh T_{\Gamma,f_{1}, f_{2}}(\x, \y), \wh T\in\{\wh T_{\text{Fisher}},\wh T_{\text{min}},\wh T_{\text{Cauchy}}\}$ are consistent against all possible alternatives under the permutation procedures.

\section{Extension to Allow for Population Unknown Function}
We note that some choices of $f_{1}$ and $f_{2}$ involve  population distribution functions, then
the oracle $\wh \mu_{\gamma,f_{1}, f_{2}}$ cannot be computed directly from the observations.
For example, $f_{j}^{\RdCov}$  includes the center-outward distribution function.
To obtain the empirical estimates $\wh \mu_{\gamma,\wh f_{1}, \wh f_{2}}$,
we need to plug in the empirical
center-outward distribution function $\wh \F_{j,\pm}$, and the corresponding
estimate of  $f_{j}^{\RdCov}$   is then given by
$
\wh f_{j}^{\RdCov}(\z_1,\ldots,\z_4)=\|\wh \F_{j,\pm}(\z_{1})-\wh \F_{j,\pm}(\z_{2})\|.
$
If $\z$ has a nonvanishing probability measure, it is implied by the Glivenko-Cantelli property of an empirical center-outward distribution function that
\beqr\label{Glivenko-Cantelli-property}
\text{$f_{j}$  is uniformly bounded and $\wh f_{j}$ almost surely converges to $f_{j}$. }
\eeqr
for $j\in\{1,2\}$.

The following proposition summarizes the asymptotic validity of the proposed tests.

{\prop{\label{proposition3}} {\rm (Generalization)}
Suppose that the conditions in (\ref{Glivenko-Cantelli-property}) and   Assumption 1 are
fulfilled.
If $\x$ and $\y$ are independent, the test statistics
$\{\wh \mu_{\gamma,\wh f_{1}, \wh f_{2}}\}_{\gamma\in\Gamma}$ are asymptotically equivalent
to their oracle versions $\{\wh \mu_{\gamma, f_{1}, f_{2}}\}_{\gamma\in\Gamma}$, that is, $w_{n, \gamma}\, \wh \mu_{\gamma,\wh f_{1}, \wh f_{2}}-w_{n, \gamma}\, \wh \mu_{\gamma, f_{1},  f_{2}}=o_{p}(1), \gamma\in\Gamma$.
}

Theorem \ref{theorem4}, Proposition \ref{proposition3} and \citet[Proposition 2.4]{Shi:Drton:Han:2022} together indicate that if we choose the
center-outward rank-based kernel functions, the resulting tests are distribution-free  within the family of non-vanishing probability measures.
Hence, obtaining p-values for rejection of the independence can be approximated via a
simulation-based procedure.

\section{Numerical Studies}\label{sec:num}
\subsection{Simulations}\label{sec:sim}

We demonstrate a wide range of applications
of our proposed independence tests, by examining their performances on both simulations and a real-world dataset.
To construct the test statistics, we consider the kernel functions $f_1=f_2=f^{\dCov}$ in Example 2.1.
For simplicity, we denote the test statistics $w_{n, \gamma}\ \wh \mu_{\gamma,f_{1}, f_{2}}(\x, \y)$ as $\wh T_{\gamma},  \gamma\in\Gamma$ and
the combined test statistics as $\{\wh T_{\text{Fisher}},\wh T_{\text{min}},\wh T_{\text{Cauchy}}\}$.
The candidate set $\Gamma=\{1,2,3,4,5,6,\infty\}$.
For $\gamma=1$, the test statistic $\wh T_{1}$ corresponds to the U-type distance covariance.
For comparison, we consider six independence testing methods, which are available
in the R software: V-type distance covariance \citep[dCov]{Szekely:Rizzo:Bakirov:2007} from the energy package, Hilbert-Schmidt independence
criteria \citep[HSIC]{Grettonetal:2008} from the dHSIC package,
Heller's graph-based test \citep[HHG]{Heller:Heller:Gorfine:2013} from the HHG package,
Multiscale Graph Correlation \citep[MGC]{Shen:Priebe:Vogelstein:2020}  from the mgc package,
projection covariance \citep[pCov]{Zhu:Xu:Li:Zhong:2017}, and center-outward rank-based distance covariance \citep[RdCov]{Shi:Drton:Han:2022}.
The nominal significance level is set to $\alpha=0.05$. We use the sample size $n=100$ and dimensions $d_1=d_2=\{5,100,200,400\}$.
The  p-values for these tests are computed through $200$ random permutations.
We report the empirical sizes and powers based on 1,000 replications.
Moreover, we also summarize the simulation results based on the kernel functions $f_1=f_2=f^{\RdCov}$ in the supplementary material,
to illustrate the effectiveness of proposed tests with distribution functions to be estimated.

To evaluate the size performances, we generate the $\x$ and $\y$ from the following two distributions: (a) multivariate normal distribution with
zero mean and banded covariance matrix $\Sigma=(\sigma_{ij})$, where $ \sigma_{ii}=1$, $\sigma_{ij}=0.5$ for $i\neq j$ and $|i-j|\leq 1$, $\sigma_{ij}=0$ for $i\neq j$ and $|i-j|> 1$.
(b) multivariate $t(3)$ distribution with independent coordinates.
The empirical sizes are charted in Table $\ref{size1}$.
It can be observed that the proposed tests as well as six competitors have good capability to maintain the type I error rates.

Since all the methods can control the size effectively, we further turn to  compare their empirical power performances.
We generate $\x$ from a multivariate uniform distribution $\calU(-1,1)^{d_1}$, unless otherwise stated. The error $\beps$ is drawn from the aforementioned normal distribution (a) and $t(3)$ distribution (b), respectively. We employ the following
five models:

(M1) $\y=\x+\kappa\beps$, where $\kappa=1.5$ with normal error  and $\kappa=0.4$ with $t(3)$ error.

(M2) $\y=\x^2+\kappa\beps$, where $\kappa=0.1$ with normal error  and $\kappa=0.05$ with $t(3)$ error.

(M3) $\x=\cos(\pi\w)+\kappa\beps$ and $\y=\sin(\pi\w)$, where $\w$ is also generated from the multivariate uniform distribution $\calU(-1,1)^{d_1}$, and $\kappa=0.5$ with normal error  and $\kappa=0.15$ with $t(3)$ error.

(M4) $\x=\w_1\cos(-\pi/4)+\w_2\sin(-\pi/4)+\kappa\beps$ and $\y=-\w_1\sin(-\pi/4)+\w_2\cos(-\pi/4)$, where $\w_1$ and $\w_2$ are independently  generated from the multivariate uniform distribution $\calU(-1,1)^{d_1}$, and $\kappa=0.05$ with both normal  and  $t(3)$ errors.

(M5) $\y=(\x^2+\kappa\beps)(W-1/2)$, where $W\sim$ Bernoulli(1, 1/2), and $\kappa=0.5$ with normal error  and $\kappa=0.1$ with $t(3)$ error.

The data generating processes are mainly adapted from the simulation settings of  \cite{Reshefetal:2011}, \cite{Simon:Tibshirani:2012} 
and \cite{Shen:Priebe:Vogelstein:2020}.  The first model (M1) includes a linear relationship, while the last four models
consist of nonmonotonic and strongly nonlinear relationships. The empirical powers for (M1)-(M5) are presented in Tables \ref{powerL}-\ref{powerP}, respectively.

From Table \ref{powerL}, we observe that all the testing methods can capture the linear relationships comparably well, except for $\RdCov$, which deteriorates rapidly as the dimension increases.
This is mainly attributed to the large positive values of $S_{1, f_{1}, f_{2}}+S_{2, f_{1}, f_{2}}-2S_{3, f_{1}, f_{2}}$, as demonstrated in Table \ref{S123}.

For quadratic relationships in Table \ref{powerQ}, all  the testing methods yield satisfactory power performances for $d_1=d_2=5$.
However, as the dimensions increase, Table \ref{S123} shows that either $S_{1, f_{1}, f_{2}}-S_{3, f_{1}, f_{2}}$ or $S_{2, f_{1}, f_{2}}-S_{3, f_{1}, f_{2}}$ takes negative values, and their absolute values are mostly comparable. This leads
to small values of $S_{1, f_{1}, f_{2}}+S_{2, f_{1}, f_{2}}-2S_{3, f_{1}, f_{2}}$.
As a consequence, dCov, HSIC, MGC, pCov, RdCov and $\wh T_1$ deteriorates sharply.
Their powers are considerably lower than those of our newly proposed  $\wh T_{\gamma}, \gamma \geq2$
and combined tests $\wh T_{\text{Fisher}}$, $\wh T_{\text{min}}$, and $\wh T_{\text{Cauchy}}$. HHG performs well, slightly trailing behind our proposed methods.

It is worth noting that the same phenomenon occurs more pronouncedly for other nonmonotonic and nonlinear relationships  in Tables \ref{powerC}-\ref{powerP}.
It appears that the powers of dCov, HSIC, MGC, pCov, RdCov and $\wh T_1$
are close to the significance level $\alpha=0.05$, due to the cancellation effects  of
$S_{1, f_{1}, f_{2}}-S_{3, f_{1}, f_{2}}$ and
$S_{2, f_{1}, f_{2}}-S_{3, f_{1}, f_{2}}$ in Table \ref{S123}.
The newly developed test statistics with $\gamma \geq2$ significantly enhance the powers of independence tests compared to the original $\wh T_1$.
The  combined tests based on $\wh T_{\text{Fisher}}$, $\wh T_{\text{min}}$, and $\wh T_{\text{Cauchy}}$ have similar powers, and
maintain consistently admirable performances across all settings.

\begin{table}[h!] 
	\begin{center}
		\caption{\label{size1} The empirical sizes of testing methods under $\alpha=0.05$.}	
				\footnotesize
		\begin{tabular}{lcccccccccc}
			\toprule
			& \multicolumn{4}{c}{Normal}&&\multicolumn{4}{c}{$t(3)$} \\\cline{2-5}\cline{7-10}
			&5&100&200&400&&5&100&200&400\\\hline
			dCov       & 0.059 & 0.061 & 0.058 & 0.058 &	  & 0.062 & 0.058 & 0.042 & 0.057 \\
			HSIC       & 0.049 & 0.060 & 0.059 & 0.059 & 	  & 0.062 & 0.061 & 0.047 & 0.055 \\
			HHG     & 0.051 & 0.045 & 0.053 & 0.060 &	  & 0.057 & 0.058 & 0.047 & 0.049  \\
			MGC&0.040& 0.057& 0.043& 0.054&&0.053 &0.044 &0.054 &0.047\\
			pCov& 0.046& 0.040& 0.059& 0.046&&0.062& 0.056 &0.050 &0.060\\
			RdCov & 0.071 & 0.052 & 0.043 & 0.044 && 0.051 & 0.058 & 0.040 & 0.057 \\
			$\wh T_{1}$          & 0.050 & 0.066 & 0.054 & 0.056 & & 0.045 & 0.048 & 0.044 & 0.059 \\
			$\wh T_{2}$         & 0.055 & 0.050 & 0.058 & 0.060 & & 0.052 & 0.051 & 0.051 & 0.046 \\
			$\wh T_{3}$         & 0.057 & 0.048 & 0.062 & 0.065 & & 0.054 & 0.054 & 0.060 & 0.048 \\
			$\wh T_{4}$         & 0.055 & 0.050 & 0.058 & 0.060 & & 0.052 & 0.051 & 0.051 & 0.046 \\
			$\wh T_{5}$         & 0.055 & 0.048 & 0.049 & 0.062 & & 0.055 & 0.051 & 0.056 & 0.049 \\
		    $\wh T_{6}$        & 0.052 & 0.050 & 0.058 & 0.061 & & 0.052 & 0.051 & 0.051 & 0.046 \\
			$\wh T_{\infty}$   & 0.055 & 0.051 & 0.050 & 0.060 & & 0.051 & 0.052 & 0.051 & 0.046 \\
			$\wh T_{\text{Fisher}}$ & 0.058 & 0.051 & 0.062 & 0.070 & & 0.061 & 0.061 & 0.058 & 0.058 \\
			$\wh T_{\text{min}}$       & 0.053 & 0.049 & 0.065 & 0.055 & & 0.055 & 0.049 & 0.054 & 0.052 \\
			$\wh T_{\text{Cauchy}}$& 0.059 & 0.056 & 0.062 & 0.066 & & 0.057 & 0.061 & 0.056 & 0.056 \\  \hline
		\end{tabular}
	\end{center}
\end{table}

\begin{table}[h!] 
	\begin{center}
		\caption{\label{powerL} The empirical powers of testing methods for (M1) under $\alpha=0.05$.}
			\footnotesize
		\begin{tabular}{lcccccccccc}
			\toprule
			& \multicolumn{4}{c}{Normal}&&\multicolumn{4}{c}{$t(3)$} \\\cline{2-5}\cline{7-10}
			&5&100&200&400&&5&100&200&400\\\hline
			dCov       & 1.000 & 1.000 & 1.000 & 1.000 && 1.000 & 1.000 & 1.000 & 1.000 \\
			HSIC       & 1.000 & 1.000 & 1.000 & 1.000 && 1.000 & 1.000 & 1.000 & 1.000 \\
			HHG     & 0.941 & 0.881 & 0.867 & 0.857 && 1.000 & 1.000 & 1.000 & 1.000 \\
			MGC&1.000 & 1.000 & 1.000 & 1.000&&1.000 & 1.000 & 1.000 & 1.000\\
			pCov& 1.000 & 1.000 & 1.000 & 1.000&&1.000 & 1.000 & 1.000 & 1.000\\
			RdCov &0.935 & 0.059 & 0.030 & 0.052 && 1.000 & 0.066 & 0.072 & 0.046 \\
				$\wh T_{1}$           & 1.000 & 1.000 & 1.000 & 1.000 & & 1.000 & 1.000 & 1.000 & 1.000 \\
				$\wh T_{2}$         & 0.915 & 0.875 & 0.878 & 0.879 & & 0.999 & 0.955 & 0.926 & 0.902 \\
				$\wh T_{3}$          & 0.993 & 0.995 & 0.990 & 0.996 & & 1.000 & 1.000 & 0.997 & 0.996 \\
				$\wh T_{4}$          & 0.951 & 0.929 & 0.933 & 0.932 & & 0.999 & 0.967 & 0.943 & 0.927 \\
				$\wh T_{5}$          & 0.976 & 0.978 & 0.973 & 0.976 & & 1.000 & 0.995 & 0.981 & 0.981 \\
			    $\wh T_{6}$       & 0.959 & 0.941 & 0.941 & 0.942 & & 0.999 & 0.975 & 0.949 & 0.934 \\
				$\wh T_{\infty}$& 0.972 & 0.967 & 0.962 & 0.963 & & 0.999 & 0.982 & 0.956 & 0.947 \\
			$\wh T_{\text{Fisher}}$ & 0.997 & 1.000 & 1.000 & 1.000 & & 1.000 & 1.000 & 1.000 & 1.000 \\
			$\wh T_{\text{min}}$   & 0.999 & 1.000 & 1.000 & 1.000 & & 1.000 & 1.000 & 1.000 & 1.000 \\
			$\wh T_{\text{Cauchy}}$ & 0.997 & 1.000 & 1.000 & 1.000 & & 1.000 & 1.000 & 1.000 & 1.000 \\ \hline
		\end{tabular}
	\end{center}
\end{table}

\begin{table}[h!] 
	\begin{center}
		\caption{\label{powerQ} The empirical powers of testing methods for (M2) under $\alpha=0.05$.}
				\footnotesize
		\begin{tabular}{lcccccccccc}
			\toprule
			& \multicolumn{4}{c}{Normal}&&\multicolumn{4}{c}{$t(3)$} \\\cline{2-5}\cline{7-10}
			&5&100&200&400&&5&100&200&400\\\hline
			dCov       & 0.999 & 0.138 & 0.128 & 0.116  & & 0.997 & 0.133 & 0.102 & 0.108 \\
			HSIC      & 1.000 & 0.182 & 0.144 & 0.129 &   & 1.000 & 0.179 & 0.125 & 0.110 \\
			HHG       & 1.000 & 0.956 & 0.938 & 0.921  &    & 1.000 & 0.955 & 0.925 & 0.933 \\
			MGC&1.000 & 0.305 & 0.187 & 0.139&&1.000 & 0.361 & 0.193 & 0.118 \\
			pCov&0.932 & 0.057 & 0.055 & 0.054&&0.967& 0.061& 0.067& 0.050 \\
			RdCov&0.602 & 0.059 & 0.041 & 0.045 &&0.617 & 0.040 & 0.041 & 0.041\\
			$\wh T_{1}$      & 0.984 & 0.065 & 0.055 & 0.055 && 0.995 & 0.065 & 0.068 & 0.052 \\
			$\wh T_{2}$     & 1.000 & 0.999 & 0.999 & 1.000 && 0.999 & 0.979 & 0.967 & 0.965 \\
			$\wh T_{3}$     & 1.000 & 0.832 & 0.874 & 0.845 && 1.000 & 0.812 & 0.789 & 0.772 \\
			$\wh T_{4}$     & 1.000 & 0.999 & 0.999 & 1.000 && 0.999 & 0.979 & 0.966 & 0.963 \\
			$\wh T_{5}$     & 1.000 & 0.974 & 0.964 & 0.970 && 1.000 & 0.941 & 0.922 & 0.921 \\
		    $\wh T_{6}$  & 1.000 & 0.999 & 0.998 & 1.000 && 0.999 & 0.978 & 0.966 & 0.962 \\
			$\wh T_{\infty}$& 1.000 & 0.998 & 0.996 & 0.996 && 0.999 & 0.977 & 0.960 & 0.960 \\
				$\wh T_{\text{Fisher}}$  & 1.000 & 1.000 & 0.998 & 1.000 && 1.000 & 0.981 & 0.964 & 0.961 \\
				$\wh T_{\text{min}}$    & 1.000 & 0.997 & 0.998 & 1.000 & & 1.000 & 0.975 & 0.959 & 0.958 \\
				$\wh T_{\text{Cauchy}}$  & 1.000 & 0.999 & 0.998 & 1.000  & & 1.000 & 0.978 & 0.963 & 0.962 \\ \hline
		\end{tabular}
	\end{center}
\end{table}

 \begin{table}[h!] 
	\begin{center}
		\caption{\label{powerC} The empirical powers of testing methods for (M3) under $\alpha=0.05$.}
				\footnotesize
		\begin{tabular}{lcccccccccc}
			\toprule
			& \multicolumn{4}{c}{Normal}&&\multicolumn{4}{c}{$t(3)$} \\\cline{2-5}\cline{7-10}
			&5&100&200&400&&5&100&200&400\\\hline
			dCov        & 0.039 & 0.033 & 0.034 & 0.029 && 0.037 & 0.026 & 0.024 & 0.026 \\
			HSIC        & 0.053 & 0.029 & 0.035 & 0.021 & & 0.078 & 0.032 & 0.026 & 0.027 \\
			HHG       & 0.397 & 0.572 & 0.507 & 0.592 && 0.879 & 0.889 & 0.862 & 0.871 \\
			MGC&0.085 & 0.052 & 0.048 & 0.053 &&0.185& 0.044& 0.031& 0.051\\
			pCov&0.045& 0.050& 0.055& 0.066&&0.056& 0.047& 0.045& 0.061\\
			RdCov&0.070 & 0.052 & 0.023 & 0.051&&0.058 & 0.051 & 0.031 & 0.050 \\
				$\wh T_{1}$      & 0.047 & 0.049 & 0.056 & 0.067 && 0.062 & 0.049 & 0.044 & 0.061 \\
				$\wh T_{2}$     & 0.957 & 0.959 & 0.949 & 0.951 && 0.998 & 0.967 & 0.949 & 0.936 \\
				$\wh T_{3}$     & 0.696 & 0.679 & 0.692 & 0.695 && 0.867 & 0.804 & 0.794 & 0.785 \\
				$\wh T_{4}$     & 0.955 & 0.958 & 0.950 & 0.944 && 0.997 & 0.967 & 0.949 & 0.934 \\
				$\wh T_{5}$     & 0.856 & 0.847 & 0.848 & 0.849 && 0.976 & 0.928 & 0.907 & 0.900 \\
			    $\wh T_{6}$  & 0.950 & 0.955 & 0.949 & 0.939 && 0.997 & 0.965 & 0.947 & 0.933 \\
				$\wh T_{\infty}$& 0.955 & 0.959 & 0.954 & 0.949 & & 0.998 & 0.964 & 0.946 & 0.939 \\
			$\wh T_{\text{Fisher}}$ & 0.940 & 0.940 & 0.932 & 0.942 && 0.996 & 0.963 & 0.947 & 0.933 \\
			$\wh T_{\text{min}}$   & 0.932 & 0.932 & 0.921 & 0.932 & & 0.996 & 0.956 & 0.939 & 0.923 \\
			$\wh T_{\text{Cauchy}}$ & 0.947 & 0.944 & 0.940 & 0.938 &  & 0.996 & 0.962 & 0.943 & 0.932 \\  \hline
		\end{tabular}
	\end{center}
\end{table}

\begin{table}[h!] 
	\begin{center}
		\caption{\label{powerD} The empirical powers of testing methods for (M4) under $\alpha=0.05$.}
				\footnotesize
		\begin{tabular}{lcccccccccc}
			\toprule
			& \multicolumn{4}{c}{Normal}&&\multicolumn{4}{c}{$t(3)$} \\\cline{2-5}\cline{7-10}
			&5&100&200&400&&5&100&200&400\\\hline
			dCov     & 0.034 & 0.026 & 0.020 & 0.016 && 0.035 & 0.021 & 0.018 & 0.018 \\
			HSIC     & 0.074 & 0.023 & 0.019 & 0.015 && 0.072 & 0.025 & 0.018 & 0.014 \\
			HHG       & 0.915 & 0.915 & 0.926 & 0.913 && 0.915 & 0.914 & 0.898 & 0.887 \\
			MGC&0.341 & 0.055 & 0.038 & 0.050&&0.312 & 0.044 & 0.047 & 0.049\\
			pCov&0.058 & 0.065 & 0.048 & 0.052&&0.052& 0.046& 0.047& 0.039\\
			RdCov&0.067 & 0.053 & 0.037 & 0.065&&0.041 & 0.059 & 0.044 & 0.042 \\
				$\wh T_{1}$      & 0.063 & 0.065 & 0.049 & 0.053 && 0.068 & 0.042 & 0.048 & 0.040 \\
				$\wh T_{2}$     & 0.991 & 0.993 & 0.992 & 0.985 && 0.983 & 0.982 & 0.981 & 0.982 \\
				$\wh T_{3}$     & 0.798 & 0.790 & 0.792 & 0.810 && 0.791 & 0.778 & 0.762 & 0.777 \\
				$\wh T_{4}$     & 0.991 & 0.993 & 0.992 & 0.985 && 0.982 & 0.982 & 0.981 & 0.981 \\
				$\wh T_{5}$     & 0.938 & 0.931 & 0.932 & 0.935 && 0.940 & 0.930 & 0.927 & 0.924 \\
			    $\wh T_{6}$  & 0.991 & 0.992 & 0.991 & 0.985 && 0.981 & 0.982 & 0.981 & 0.982 \\
				$\wh T_{\infty}$& 0.990 & 0.992 & 0.989 & 0.988 && 0.979 & 0.981 & 0.984 & 0.981\\
			$\wh T_{\text{Fisher}}$  & 0.993 & 0.989 & 0.986 & 0.981 && 0.977 & 0.979 & 0.975 & 0.978 \\
			$\wh T_{\text{min}}$   & 0.988 & 0.985 & 0.972 & 0.974 && 0.974 & 0.971 & 0.965 & 0.969 \\
			$\wh T_{\text{Cauchy}}$ & 0.992 & 0.989 & 0.984 & 0.984& & 0.979 & 0.981 & 0.976 & 0.979  \\\hline
		\end{tabular}
	\end{center}
\end{table}

\begin{table}[h!] 
	\begin{center}
		\caption{\label{powerP} The empirical powers of testing methods for (M5) under $\alpha=0.05$.}
				\footnotesize
		\begin{tabular}{lcccccccccc}
			\toprule
			& \multicolumn{4}{c}{Normal}&&\multicolumn{4}{c}{$t(3)$} \\\cline{2-5}\cline{7-10}
			&5&100&200&400&&5&100&200&400\\\hline
			dCov       & 0.114 & 0.094 & 0.093 & 0.099 &    & 0.292 & 0.169 & 0.173 & 0.171  \\
			HSIC      & 0.161 & 0.096 & 0.095 & 0.103 && 0.666 & 0.183 & 0.177 & 0.179 \\
			HHG       & 0.690 & 0.754 & 0.761 & 0.747 && 1.000 & 1.000 & 1.000 & 1.000 \\
			MGC&0.091 & 0.053 & 0.067 & 0.052&&0.786 & 0.066 & 0.070 & 0.072 \\
			pCov&0.065& 0.035 &0.048& 0.058&&0.118& 0.058& 0.054& 0.045\\
			RdCov&0.082 & 0.045 & 0.042 & 0.050&& 0.125 & 0.049 & 0.057 & 0.035\\
			$\wh T_{1}$            & 0.063 & 0.035 & 0.048 & 0.059 & & 0.141 & 0.060 & 0.053 & 0.045 \\
			$\wh T_{2}$     & 0.940 & 0.933 & 0.910 & 0.916 & & 0.997 & 0.991 & 0.981 & 0.975 \\
			$\wh T_{3}$     & 0.688 & 0.664 & 0.646 & 0.683 & & 0.951 & 0.934 & 0.905 & 0.896 \\
			$\wh T_{4}$     & 0.938 & 0.933 & 0.911 & 0.916 & & 0.997 & 0.991 & 0.981 & 0.975 \\
			$\wh T_{5}$     & 0.826 & 0.811 & 0.803 & 0.829 & & 0.993 & 0.984 & 0.968 & 0.957 \\
		    $\wh T_{6}$  & 0.938 & 0.934 & 0.909 & 0.915 & & 0.997 & 0.991 & 0.981 & 0.975 \\
			$\wh T_{\infty}$& 0.919 & 0.914 & 0.893 & 0.901 & & 0.997 & 0.991 & 0.981 & 0.971 \\
			$\wh T_{\text{Fisher}}$ & 0.916 & 0.916 & 0.898 & 0.906 & & 0.998 & 0.988 & 0.980 & 0.974 \\
			$\wh T_{\text{min}}$   & 0.893 & 0.886 & 0.864 & 0.872 & & 0.998 & 0.986 & 0.976 & 0.969 \\
			$\wh T_{\text{Cauchy}}$& 0.918 & 0.917 & 0.898 & 0.907 & & 0.998 & 0.989 & 0.980 & 0.972 \\ \hline
		\end{tabular}
	\end{center}
\end{table}


\subsection{A real data analysis}
In this section, we provide an empirical analysis of the proposed methods using a gene expression microarray dataset, which is
collected from twelve-week-old male rats  \citep{scheetz2006regulation}.
It contains tissue samples harvested from the eyes of 120 rats, with a total of 18,976 probe sets exhibiting sufficient expression signals.
We focus on the gene TRIM32 at probe 1389163$\_$at, which has been found to  cause the Bardet-Biedl syndrome \citep{chiang2006homozygosity}.
For the remaining genes, we select 300 probe sets that have the largest variances.
  We are interested in whether these selected genes
  are associated with the gene TRIM32.

\begin{figure}[htp!]
	\centerline{
		\begin{tabular}{ccc}
			\psfig{figure=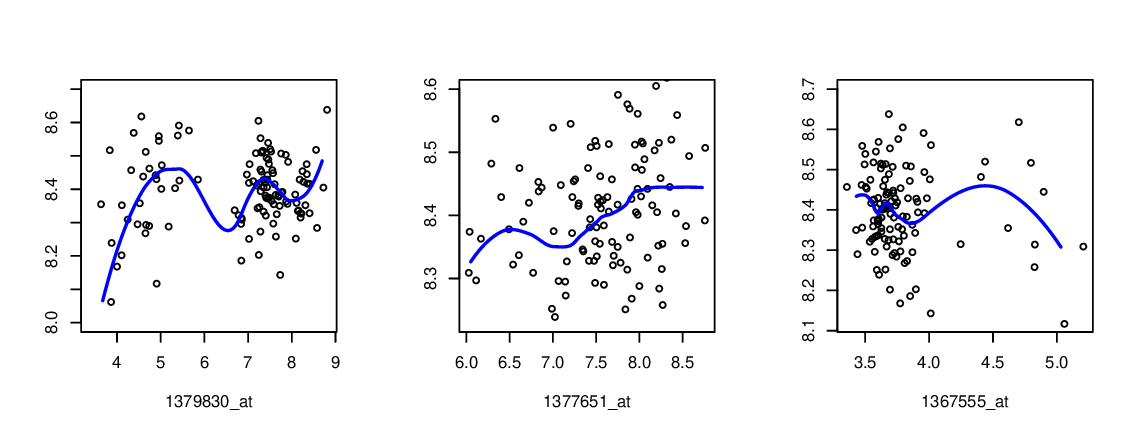,width=6.5in,height=2.5in,angle=0}
	\end{tabular}}
	\caption{\label{fig:gen}  The scatter plot of selected genes expression levels versus   TRIM32.
	}
\end{figure}

  We apply the proposed tests with  $f_1=f_2=f^{\dCov}$ to analyze this dataset, compared to the six competitors mentioned in the simulation section.
 To evaluate their power performances, we
  randomly select subsets of size $n=45$, 60, 75 and 90 from the whole data set.
 We report the empirical powers of  the testing methods based on 200 repetitions in Table \ref{real1}.
 It can be seen that the newly proposed tests based on $\wh T_\gamma$ with $\gamma\geq2$ and combined test  statistics $\wh T_{\text{Fisher}}$, $\wh T_{\text{min}}$, and $\wh T_{\text{Cauchy}}$
 generally outperform the competitors dCov, HSIC, HHG, MGC, pCov, RdCov and $\wh T_1$.
To  get an insight into the power performances,
 we further calculate the values of  $S_{1, f_{1}, f_{2}}-S_{3, f_{1}, f_{2}}$ and $S_{2, f_{1}, f_{2}}-S_{3, f_{1}, f_{2}}$, which are charted in Table \ref{real2}.
 We can observe that $S_{1, f_{1}, f_{2}}-S_{3, f_{1}, f_{2}}$ and $S_{2, f_{1}, f_{2}}-S_{3, f_{1}, f_{2}}$ offset
 each other's effects, leading to small values of $S_{1, f_{1}, f_{2}}+S_{2, f_{1}, f_{2}}-2S_{3, f_{1}, f_{2}}$.
 This may partially explain the reasons for  low powers of the existing methods.
 Since genes at 1379830$\_$at, 1377651$\_$at and 1367555$\_$at rank at the top three for most of our proposed tests, we 
 provide a scatter plot of them
 versus TRIM32 to visualize the relationships.
Figure \ref{fig:gen} shows that there exist nonlinear relationships between the selected genes and TRIM32.

\begin{table}[h!] 
	\begin{center}
		\caption{\label{real1} The empirical powers of testing methods under $\alpha=0.05$.}	
	\footnotesize
	\begin{tabular}{lcccclccccc}
		\toprule
		& \multicolumn{4}{c}{Sample size}&&\multicolumn{4}{c}{Sample size} \\\cline{2-5}\cline{7-10}
	Method	&45&60&75&90&Method&45&60&75&90\\\hline
		dCov       & 0.355 & 0.470 & 0.700 & 0.925&          $\wh T_{3}$         	& 0.455 & 0.655 & 0.865 & 0.990 \\
		HSIC       &   0.235 & 0.320 & 0.475 & 0.625& $\wh T_{4}$         	& 0.470 & 0.650 & 0.805 & 0.925 \\
		HHG       & 0.195 & 0.250 & 0.395 & 0.560 & $\wh T_{5}$         	& 0.450 & 0.690 & 0.870 & 0.985 \\
		MGC    & 0.245 & 0.375 & 0.400 & 0.445&$\wh T_{6}$         	& 0.470 & 0.670 & 0.795 & 0.925 \\
		pCov    &0.330& 0.460& 0.650& 0.870&$\wh T_{\infty}$   	& 0.500 & 0.715 & 0.845 & 0.955 \\
		RdCov&0.090 &0.070 &0.070 &0.090&$\wh T_{\text{Fisher}}$	& 0.495 & 0.720 & 0.875 & 0.990 \\
		$\wh T_{1}$& 0.280 & 0.345 & 0.505 & 0.725&        $\wh T_{\text{min}}$  	& 0.455 & 0.655 & 0.790 & 0.965 \\
		$\wh T_{2}$    & 0.460 & 0.630 & 0.780 & 0.915&    $\wh T_{\text{Cauchy}}	$& 0.510 & 0.715 & 0.865 & 0.995 \\\hline

	\end{tabular}
\end{center}
\end{table}

\begin{table}[h!] 
	\begin{center}
		\caption{\label{real2} $S_{1}-S_{3}$, $S_{2}-S_{3}$ and their summation with different sample sizes.}	
	\footnotesize
	\begin{tabular}{ccccc}
		\toprule
		& \multicolumn{4}{c}{Sample size} \\\cline{2-5}
		&45&60&75&90\\\hline
	
		$S_{1}-S_{3}$ & 0.069 & 0.067 & 0.071 & 0.068 \\
		$S_{2}-S_{3}$ & -0.060 & -0.057 & -0.060 & -0.057 \\
		$S_{1}+S_{2}-2S_{3}$ & 0.009 & 0.010 & 0.011 & 0.011 \\ \hline

	\end{tabular}
\end{center}
\end{table}

\section{Discussion}
\label{sec:conc}
In this article, we find that the multivariate independence testing problem can be equivalently transformed into testing whether
the bivariate mean vectors $(S_{1, f_{1}, f_{2}}, S_{3, f_{1}, f_{2}})$ and $(S_{3, f_{1}, f_{2}}, S_{2, f_{1}, f_{2}})$ are the same.
Based on this transformation, we further notice that the cancellation of $S_{1, f_{1}, f_{2}}- S_{3, f_{1}, f_{2}}$ and $S_{2, f_{1}, f_{2}}- S_{3, f_{1}, f_{2}}$ may ruin the powers
of some existing tests. 
These observations motivate us to propose new tests to guarantee power performances.
We study the asymptotic behaviors  of new tests
 thoroughly
under the null hypothesis of independence and the alternatives of general dependence.
Since we face to a wide range of dependence patterns in practice, we further develop combined probability tests  to
maintain high powers. 
Our numerical results echo the theoretical findings under various types of strongly nonmonotonic and nonlinear relationships.
The concept to boost the testing power can also be extended to test mutual independence for a single multivariate vector, and to test conditional independence.


%

%

\bibliographystyle{refstyle}

\bibliography{reference}

\begin{thebibliography}{37}
\newcommand{\enquote}[1]{``#1''}
\providecommand{\natexlab}[1]{#1}
\expandafter\ifx\csname urlstyle\endcsname\relax
  \providecommand{\doi}[1]{doi:\discretionary{}{}{}#1}\else
  \providecommand{\doi}{doi:\discretionary{}{}{}\begingroup
  \urlstyle{rm}\Url}\fi

\bibitem[{Azadkia and Chatterjee(2021)}]{Azadkia:Chatterjee:2021}
Azadkia, M. and Chatterjee, S. (2021).
\newblock \enquote{A simple measure of conditional dependence.}
\newblock \emph{The Annals of Statistics}, \textbf{49}, 3070--3102.

\bibitem[{Bergsma and Dassios(2014)}]{Bergsma:Dassios:2014}
Bergsma, W. and Dassios, A. (2014).
\newblock \enquote{A consistent test of independence based on a sign covariance
  related to kendall's tau.}
\newblock \emph{Bernoulli}, \textbf{20}, 1006--1028.

\bibitem[{Berrett and Samworth(2019)}]{Berrett:Samworth:2019}
Berrett, T. and Samworth, R. (2019).
\newblock \enquote{Nonparametric independence testing via mutual information.}
\newblock \emph{Biometrika}, \textbf{106}, 547--566.

\bibitem[{Blum et~al.(1961)Blum, Kiefer, and
  Rosenblatt}]{Blum:Kiefer:Rosenblatt:1961}
Blum, J., Kiefer, J., and Rosenblatt, M. (1961).
\newblock \enquote{Distribution free tests of independence based on the sample
  distribution function.}
\newblock \emph{The Annals of Mathematical Statistics}, \textbf{32}, 485--498.

\bibitem[{Chiang et~al.(2006)Chiang, Beck, Yen, Tayeh, Scheetz, Swiderski,
  Nishimura, Braun, Kim, Huang et~al.}]{chiang2006homozygosity}
Chiang, A.P., Beck, J.S., Yen, H.J., Tayeh, M.K., Scheetz, T.E., Swiderski,
  R.E., Nishimura, D.Y., Braun, T.A., Kim, K.Y.A., Huang, J., et~al. (2006).
\newblock \enquote{Homozygosity mapping with snp arrays identifies trim32, an
  e3 ubiquitin ligase, as a bardet--biedl syndrome gene (bbs11).}
\newblock \emph{Proceedings of the National Academy of Sciences},
  \textbf{103(16)}, 6287--6292.

\bibitem[{Deb and Sen(2023)}]{Deb:Sen:2023}
Deb, N. and Sen, B. (2023).
\newblock \enquote{Multivariate rank-based distribution-free nonparametric
  testing using measure transportation.}
\newblock \emph{To appear in Journal of the American Statistical Association}.

\bibitem[{Dhar et~al.(2016)Dhar, Dassios, and
  Bergsma}]{Dhar:Dassios:Bergsma:2016}
Dhar, S.S., Dassios, A., and Bergsma, W. (2016).
\newblock \enquote{A study of the power and robustness of a new test for
  independence against contiguous alternatives.}
\newblock \emph{Electronic Journal of Statistics}, \textbf{10}, 330--351.

\bibitem[{Fisher(1925)}]{Fisher:1925}
Fisher, R.A. (1925).
\newblock \emph{Statistical Methods for Research Workers}.
\newblock Vol. 1, Edinburgh by Oliver and Boyd.

\bibitem[{Gorsky and Ma(2022)}]{Gorsky:Ma:2022}
Gorsky, S. and Ma, L. (2022).
\newblock \enquote{Multi-scale fisher's independence test for multivariate
  dependence (with discussions).}
\newblock \emph{Biometrika}, \textbf{109}, 569--587.

\bibitem[{Gretton et~al.(2008)Gretton, Fukumizu, Teo, Song, Sch\"olkopf, and
  Smola}]{Grettonetal:2008}
Gretton, A., Fukumizu, K., Teo, C., Song, L., Sch\"olkopf, B., and Smola, A.
  (2008).
\newblock \enquote{A kernel statistical test of independence.}
\newblock \emph{In Advances in Neural Information Processing Systems}, pages
  585--592.

\bibitem[{Hallin(2017)}]{Hallin:2017}
Hallin, M. (2017).
\newblock \enquote{On distribution and quantile functions, ranks and signs in
  $r^d$: A measure transportation approach.}
\newblock \emph{available at
  https://ideas.repec.org/p/eca/wpaper/2013-258262.html.}

\bibitem[{He et~al.(2021)He, Xu, Wu, and Pan}]{He:Xu:Wu:Pan:2021}
He, Y., Xu, G., Wu, C., and Pan, W. (2021).
\newblock \enquote{Asymptotically independent u-statistics in high-dimensional
  testing.}
\newblock \emph{The Annals of Statistics}, \textbf{49}, 154--181.

\bibitem[{Heller and Heller(2016)}]{Heller:Heller:2016}
Heller, R. and Heller, Y. (2016).
\newblock \enquote{Multivariate tests of association based on univariate
  tests.}
\newblock \emph{Advances in Neural Information Processing Systems 29 (NIPS
  2016)}.

\bibitem[{Heller et~al.(2013)Heller, Heller, and
  Gorfine}]{Heller:Heller:Gorfine:2013}
Heller, R., Heller, Y., and Gorfine, M. (2013).
\newblock \enquote{A consistent multivariate test of association based on ranks
  of distances.}
\newblock \emph{Biometrika}, \textbf{100}, 503--510.

\bibitem[{Heller et~al.(2016)Heller, Heller, Kaufman, Brill, and
  Gorfine}]{heller2016consistent}
Heller, R., Heller, Y., Kaufman, S., Brill, B., and Gorfine, M. (2016).
\newblock \enquote{Consistent distribution-free k-sample and independence tests
  for univariate random variables.}
\newblock \emph{The Journal of Machine Learning Research}, \textbf{17(1)},
  978--1031.

\bibitem[{Hoeffding(1948b)}]{Hoeffding:1948}
Hoeffding, W. (1948b).
\newblock \enquote{A non-parametric test of independence.}
\newblock \emph{The Annals of Mathematical Statistics}, \textbf{19}, 546--557.

\bibitem[{Kendall(1938)}]{Kendall:1938}
Kendall, M. (1938).
\newblock \enquote{A new measure of rank correlation.}
\newblock \emph{Biometrika}, \textbf{30}, 81--93.

\bibitem[{Kim et~al.(2020a)Kim, Balakrishnan, and
  Wasserman}]{Kim:Balakrishnan:Wasserman:2020a}
Kim, I., Balakrishnan, S., and Wasserman, L. (2020a).
\newblock \enquote{Robust multivariate nonparametric tests via projection
  averaging.}
\newblock \emph{The Annals of Statistics}, \textbf{48}, 3417--3441.

\bibitem[{Korolyuk and Borovskich(2013)}]{Korolyuk:Borovskich:2013}
Korolyuk, V.S. and Borovskich, Y.V. (2013).
\newblock \emph{Theory of U-statistics}.
\newblock Springer Science \& Business Media.

\bibitem[{Lee et~al.(2022)Lee, Zhang, and Kosorok}]{Lee:Zhang:Kosorok:2022}
Lee, D., Zhang, K., and Kosorok, M.R. (2022).
\newblock \enquote{Testing independence with the binary expansion randomized
  ensemble test.}
\newblock \emph{To appear in Statistica Sinica}.

\bibitem[{Littell and Folks(1971)}]{Littell:Folks:1971}
Littell, R.C. and Folks, J.L. (1971).
\newblock \enquote{Asymptotic optimality of fisher's method of combining
  independent tests.}
\newblock \emph{Journal of the American Statistical Association}, \textbf{66},
  802--806.

\bibitem[{Littell and Folks(1973)}]{Littell:Folks:1973}
Littell, R.C. and Folks, J.L. (1973).
\newblock \enquote{Asymptotic optimality of fisher's method of combining
  independent tests ii.}
\newblock \emph{Journal of the American Statistical Association}, \textbf{68},
  193--194.

\bibitem[{Liu and Xie(2020)}]{Liu:Xie:2020}
Liu, Y. and Xie, J. (2020).
\newblock \enquote{Cauchy combination test: a powerful test with analytic
  p-value calculation under arbitrary dependency structures.}
\newblock \emph{Journal of the American Statistical Association}, \textbf{115},
  393--402.

\bibitem[{Moon and Chen(2020)}]{Moon:Chen:2020}
Moon, H. and Chen, K. (2020).
\newblock \enquote{Interpoint-ranking sign covariance for test of
  independence.}
\newblock \emph{Biometrika}, \textbf{103}, 1--14.

\bibitem[{Pan et~al.(2020)Pan, Wang, Zhang, Zhu, and Zhu}]{Panetal:2020}
Pan, W., Wang, X., Zhang, H., Zhu, H., and Zhu, J. (2020).
\newblock \enquote{Ball covariance: a generic measure of dependence in banach
  space.}
\newblock \emph{Journal of the American Statistical Association}, \textbf{115},
  307--317.

\bibitem[{Pearson(1895)}]{Pearson:1895}
Pearson, K. (1895).
\newblock \enquote{Notes on regression and inheritance in the case of two
  parents.}
\newblock \emph{Proceedings of the Royal Society of London}, \textbf{58},
  240--242.

\bibitem[{Reshef et~al.(2011)Reshef, Reshef, Finucane, Grossman, McVean,
  Turnbaugh, Lander, Mitzenmacher, and Sabeti}]{Reshefetal:2011}
Reshef, D., Reshef, Y., Finucane, H., Grossman, S., McVean, G., Turnbaugh, P.,
  Lander, E., Mitzenmacher, M., and Sabeti, P. (2011).
\newblock \enquote{Detecting novel associations in large data sets.}
\newblock \emph{Science}, \textbf{334}, 1518--1524.

\bibitem[{Scheetz et~al.(2006)Scheetz, Kim, Swiderski, Philp, Braun, Knudtson,
  Dorrance, DiBona, Huang, Casavant et~al.}]{scheetz2006regulation}
Scheetz, T.E., Kim, K.Y.A., Swiderski, R.E., Philp, A.R., Braun, T.A.,
  Knudtson, K.L., Dorrance, A.M., DiBona, G.F., Huang, J., Casavant, T.L.,
  et~al. (2006).
\newblock \enquote{Regulation of gene expression in the mammalian eye and its
  relevance to eye disease.}
\newblock \emph{Proceedings of the National Academy of Sciences},
  \textbf{103(39)}, 14429--14434.

\bibitem[{Shen et~al.(2020)Shen, Priebe, and
  Vogelstein}]{Shen:Priebe:Vogelstein:2020}
Shen, C., Priebe, C., and Vogelstein, J. (2020).
\newblock \enquote{From distance correlation to multiscale graph correlation.}
\newblock \emph{Journal of the American Statistical Association}, \textbf{115},
  280--291.

\bibitem[{Shi et~al.(2022{\natexlab{a}})Shi, Drton, and
  Han}]{Shi:Drton:Han:2022}
Shi, H., Drton, M., and Han, F. (2022{\natexlab{a}}).
\newblock \enquote{Distribution-free consistent independence tests via
  center-outward ranks and signs.}
\newblock \emph{Journal of the American Statistical Association}, \textbf{117},
  395--410.

\bibitem[{Shi et~al.(2022{\natexlab{b}})Shi, Hallin, Drton, and
  Han}]{Shi:Hallin:Drton:Han:2022}
Shi, H., Hallin, M., Drton, M., and Han, F. (2022{\natexlab{b}}).
\newblock \enquote{On universally consistent and fully distribution-free rank
  tests of vector independence.}
\newblock \emph{The Annals of Statistics}, \textbf{50}, 1933--1959.

\bibitem[{Simon and Tibshirani(2012)}]{Simon:Tibshirani:2012}
Simon, N. and Tibshirani, R. (2012).
\newblock \enquote{Comment on ``detecting novel associations in large data
  sets".}
\newblock \emph{arXiv no. 1401.7645}.

\bibitem[{Spearman(1904)}]{Spearman:1904}
Spearman, C. (1904).
\newblock \enquote{The proof and measurement of association between two
  things.}
\newblock \emph{The American Journal of Psychology}, \textbf{15}, 72--101.

\bibitem[{Sz\'ekely et~al.(2007)Sz\'ekely, Rizzo, and
  Bakirov}]{Szekely:Rizzo:Bakirov:2007}
Sz\'ekely, G., Rizzo, M., and Bakirov, N. (2007).
\newblock \enquote{Measuring and testing dependence by correlation of
  distances.}
\newblock \emph{The Annals of Statistics}, \textbf{35}, 2769--2794.

\bibitem[{Weihs et~al.(2018)Weihs, Drton, and
  Meinshausen}]{Weihs:Drton:Meinshausen:2018}
Weihs, L., Drton, M., and Meinshausen, N. (2018).
\newblock \enquote{Symmetric rank covariances: a generalized framework for
  nonparametric measures of dependence.}
\newblock \emph{Biometrika}, \textbf{105}, 547--562.

\bibitem[{Xu et~al.(2016)Xu, Lin, Wei, and Pan}]{Xu:Lin:Wei:Pan:2016}
Xu, G., Lin, L., Wei, P., and Pan, W. (2016).
\newblock \enquote{An adaptive two-sample test for high-dimensional means.}
\newblock \emph{Biometrika}, \textbf{103}, 609--624.

\bibitem[{Zhu et~al.(2017)Zhu, Xu, Li, and Zhong}]{Zhu:Xu:Li:Zhong:2017}
Zhu, L., Xu, K., Li, R., and Zhong, W. (2017).
\newblock \enquote{Projection correlation between two random vectors.}
\newblock \emph{Biometrika}, \textbf{104}, 829--843.

\end{thebibliography}
\end{document}